# An Experimental Double-Integrating Sphere Spectrophotometer for In Vitro Optical Analysis of Blood and Tissue Samples, Including Examples of Analyte Measurement Results

Daniel J. Schwab[a], Clifton R. Haider[a], Gary S. Delp[a1] Stefan K. Grebe[b] and Barry K. Gilbert[a]
[a]Mayo Clinic, Special Purpose Processor Development Group, Rochester MN 55905
[b]Mayo Clinic, Division of Laboratory Medicine and Pathology, Rochester MN 55905

## ABSTRACT

Data-driven science requires data to drive it. Being able to make accurate and precise measurement of biomaterials in the body means that medical assessments can be more accurate. There are differences between how blood absorbs and how it reflects light. The Mayo Clinic's Double-Integrating Sphere Spectrophotometer (MDISS) is an automated measurement device that detects both scattered and direct energy as it passes through a sample in a holder. It can make over 1,200 evenly spaced color measurements from the very deep purple (300-nm) through the visible light spectrum into the near infrared (2800-nm). The MDISS samples measured have been also measured by commercial laboratory equipment. The MDISS measurements are as accurate and more precise than those devices now in use.

With so many measurements to be made during the time that the sample remains undegraded, mechanical and data collection automation was required. The MDISS sample holders include different thicknesses, versions that can operate at high pressure (such as divers may experience), and versions that can pump and rotate the measured material to maintain consistency of measurement. Although the data obtained are preliminary, they have potential to guide the design of new devices for more accurate assessments. There is an extensive "lessons learned" section.

New wearable medical measurement devices can provide important real-time measurements. As an example, firefighters could be monitored for excessive carbon-monoxide buildup in time to react and save lives. The design of many such devices can benefit from detailed data of a biological material's optical absorption and reflection over a wide wavelength range. The MDISS was designed to make these measurements. Results are presented for blood oxygen ($O2_{SAT}$), carbon dioxide ($CO_2$), and carbon-monoxide (CO) at various dilutions. After obtaining donors' consent as approved by the Mayo Clinic institutional review board (IRB), samples were prepared and measured by current commercial instruments, followed by detailed measurements with the MDISS. The measured data are presented in graphs that encourage visual comparison of the various levels and ratios of these materials. Wavelengths of light can be selected for new devices based on data of this quality. The measurements indicate the potential for clinical utility of the MDISS system. ***Nota bena***: The example results presented do not represent new foundational values; rather, the data are illustrative of what can be measured.

[1] Corresponding Author: Delp.Gary at Mayo.edu or Gary.Delp at SilverLoon.Systems

**Significance:** Accurate and precise measurement of *in-vivo* biomaterials improves the tools for medical assessment. Detailed spectral response data of bio-analytes are a critical enabler for designing body-worn units to collect these measurements. A research instrument to make these measurements is described. The data collected show the potential to distinguish between distinct bio-analytes and concentrations, *e.g.,* carbon-monoxide buildup in firefighters could be monitored in real time.

**Aim:** Present the design, test, and available choices for an automated instrument for rapid measurement of multispectral, diffuse, and direct path response of bio-analytes. Provide a basis for further development. Distinguish varied concentrations and ratios of multiple hemoglobin species and related analytes.

**Approach**: The base design included a narrow-band laser, wavelength-modulated with an optical parametric oscillator (OPO), illuminating a sample sandwiched between double integrating spheres. Control and measurement refinements, automation, and calibrated sample holding mechanisms were developed and tested. Using the MDISS, measurements were compiled at 2 nm steps over a spectral range of 200–2800-nm. Energy data were collected at each step: the input energy, diffuse-reflected (DR), diffuse transmitted (DT), and unscattered-transmitted (UT) energy over stepped concentrations of multiple hemoglobin species.

**Results**: The Mayo Clinic's Double-Integrating Sphere Spectrophotometer (MDISS) was designed, fabricated, and used. Repeatable, precision data can be collected at 2 nm steps over a spectral range of 200–2800-nm. Features include rapid high throughput automated measurement capability, simultaneous acquisition of input, DR, DT, and UT energy, and the flexibility to accommodate a wide variety of sample holder types and volumes. Feature tradeoffs and potential future extensions are described. Data are presented in graphical form from samples whose oxygen concentration had been decreased with the desaturation gases carbon dioxide (CO2) and carbon monoxide (CO). A discussion is presented of the specific results that might be used to distinguish among analytes.

**Conclusions:** The measurements show potential for clinical viability of the MDISS system. The wide-range, narrow-bandwidth measurements can help to identify the optimal wavelengths to detect a specific set of bio-analytes. While these results are very preliminary data, they may justify further exploration.

## 1. BACKGROUND

The ability to monitor clinically important analytes using a variety of sensors, especially unobtrusive body-worn units, is critical for the next generations of in-home, ambulatory, and in-clinic care for an aging population. Our companion paper describes a research timeline leading to the need for and the utility of the MDISS [1]. Based on research described elsewhere [2, 3, 4], and in two of our patents [5,6], it became apparent that such monitoring could be accomplished. These publications teach the concept that the blood and tissue concentrations of analytes can be monitored quantitatively using a small number of narrowband light sources over a broad range of wavelengths. More specifically, we found that those wavelengths can be *optimally* selected using the mathematics described in patent [6], *if and only* if accurate, high wavelength resolution, measured fingerprints of clinically relevant analytes over a wide range of optical frequencies (e.g., 190 nm to 3000 nm or more) are available. With the availability of that type of comprehensive optical information, it would be possible to design autonomous body-worn analyte measurement units with significantly improved capabilities compared to those presently available. Discussions describing the past and present state of the art of such units appear in companion manuscripts to this one [1, 7].

### 1.1. Related Research in Spectrophotometric Systems

Researchers have addressed these requirements as reported in, *e.g.*, [8-12]. The Rehman paper [8] provides a review of the field with 158 references. A representative subset of these manuscripts that include implementation details are used here to compare alternatives to our system. All used double integrating spheres, but beyond that, there were several light sources, differences in sensors and sensor placement, different wavelength ranges, and other contrasting characteristics. We note the features unique to each implementation. Zamora-Rohas *et al*. [9] used a supercontinuum laser source with a bandwidth of 460–2400 nm, at a power level of 4W, with a source wavelength range of 700–2400 nm; the data that the Zamora-Rohas team presented was from 1150–2250 nm. They implemented high-sensitivity thermoelectrically cooled InGaAs photodiode detectors. The manuscript did not mention any operational automation features. The implementation described in Roggan *et al*. [10] was the most like the Mayo system: It used a sample circulation pump, a lamp-and-monochromator light source, with an 8-nm FWHM wavelength step size from 400-1100 nm and a 16-nm FWHM step size from 1140–2500 nm, as well as dual photodiode detectors for the different wavelength spans; there was no mention of system automation. Lemaillet *et al*. [11, 12] implemented a double-integrating sphere system with a HeNe laser generating two wavelengths, 543 and 642 nm, but with no apparent ability to generate other wavelengths. The beam was continuous, with no apparent ability to make very short-duration measurements. Three photodiode detectors were employed; measurements appeared to be restricted to straight-through transmission and reflectance, from the incident beam, and from the reflectance and transmittance spheres. There was no indication of an automation overlay.

### 1.2. Features and Capabilities Apparently Unique to the MDISS

As both literature and manufacturer searches revealed no commercially available instruments that could provide the needed wavelength-breadth, -resolution, and other features for such "optical fingerprints" for each of many analytes, we chose to develop the MDISS.

The key features and capabilities that differentiate the MDISS from the units described above include: a broad spectral range, i.e., 192–2750 nm compared to 700–2400 nm or 400–2500 nm; narrower line widths; 2/4 nm FWHM seeded/unseeded, compared to the 8 and 10 nm FWHM of others; smaller wavelength step sizes, ≤1 nm versus 8 nm or 10 nm steps for the other systems; likely the ability to sustain higher throughput due to automation (no automation details were presented in the referenced manuscripts); a more flexible sample holder approach (versus one fixed cell size described in each of the other manuscripts), including different path lengths, cell diameters and cell volumes, and the ability to install cuvettes; implementation of a rotation/rocking mechanism to ensure continuous mixing and suspension of particulates; a through-cell circulation system for use when needed; a flexible test environment (only minimal details were presented regarding the test environment in the referenced manuscripts); and finally, the ability to rapidly obtain a comprehensive set of measurements, thereby avoiding degradation of time-sensitive samples provided by our clinical collaborators.

### 1.3. Notional Capabilities

The general understanding was that if N analytes are to be detected quantitatively, N independent measurements at N independent wavelengths are needed. Recent work in this laboratory suggested that, if the interactions of the light at each given wavelength are recorded and measured in a multi-modality manner (e.g., forward scattered, backward scattered, unscattered, etc.), each of these modalities would provide an additional, partially independent, measure that can be used to solve the concentration equations. Thus, a smaller number of wavelengths are needed to characterize an analyte, if they are optimally selected.

A generic depiction of the utility of specific wavelength selection appears in Figure 1. The wide gray line represents a statistically normal curve of a population; typically, measured data points fall within the gray line. Any deviation outside of this wide line may represent a physiologically anomalous analyte or otherwise abnormal clinical finding. This band represents any one or a combination of, the forward scattered, backward scattered, and/or unscattered energy values. As can be observed in this figure, different anomalies may appear in different subregions in the overall wavelength span, in return requiring different and specific wavelengths, as well as different sensitivities (to determine the magnitude of the deviation from normal) for optimal detection of the anomaly. This depiction underscores the requirement for a broad wavelength span, high wavelength accuracy, precise wavelength selection, high sensitivity measurement system capable of producing repeatable baseline data that can be incorporated into wavelength selection for small autonomous (*i.e.*, untethered) body-worn continuous physiological monitoring units. It was our subsequent intended development of such portable units, planned as a next-generation implementation of prior body-worn systems previously conducted here [7] which drove these investigations.

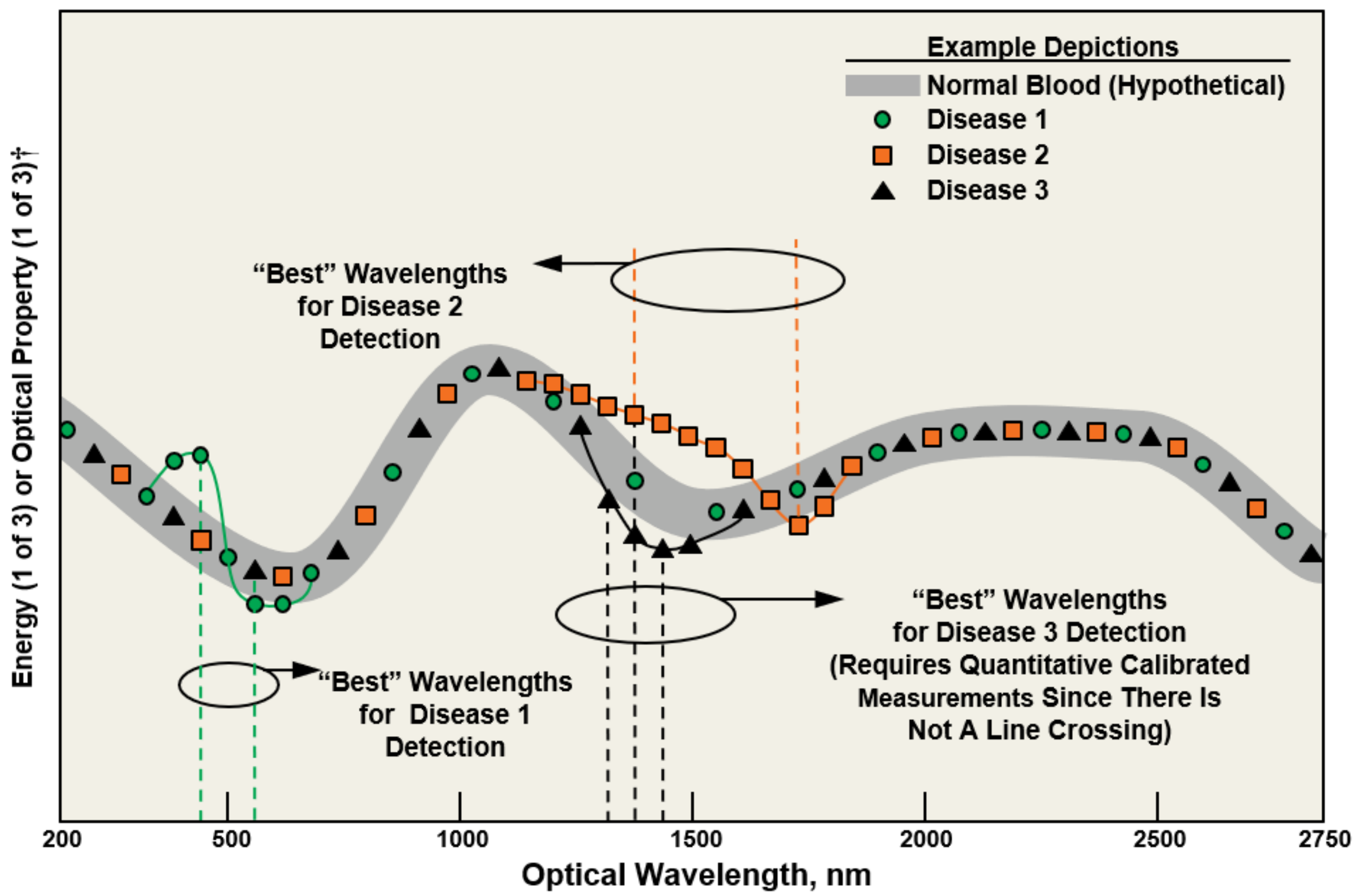

**Figure 1** MDISS measurement curves depiction. Small deviations from the "normal" curve determine which optical wavelengths are best for detecting selected physiological or biochemical anomalies. †Energy – any of three parameters: diffuse reflective, diffuse transmitted, or unscattered transmitted; Optical properties – any of three parameters: absorption coefficient, scattering coefficient, anisotropy (44471)[2]

The organization of the rest of this paper comprises: section 2 is a detailed description of the MDISS system; section 3 details a broad selection of lessons learned with suggestions to future instrument and measurement teams; section 4 contains a representative sample of the types of measurements and distinguishing measurements using the MDISS; and section 5 contains a summary and our conclusions.

## 2. THE EXPERIMENTAL SPECTROPHOTOMETER – MDISS

The information described above led to the recognition of the requirement for a spectrophotometer with an unusual set of functional characteristics that led us to design and fabricate a unit "from scratch", which we refer to as the Mayo Double Integrating Sphere Spectrophotometer (MDISS), to incorporate the features described above and to extend the results of several prior research efforts noted above [8–12]. We recognized that the MDISS, with its broadband capability to detect anomalies in blood and tissue samples, could find direct application in clinical laboratories providing a standalone direct measurement capability. Ultimately, a combination of the pump-based laser source and an OPO was selected as the best fundamental component and acquired from Continuum Inc.   Consideration was given to providers of candidate building blocks who were willing to incorporate additional features such as test automation.  Through Continuum's continued willingness to provide one-off technical support, many extensions were incorporated over time to complete the MDISS implementation.

[2] The numbers trailing the figure captions denote the figure's location in the Mayo Clinic SPPDG image archive.

### 2.1. MDISS System Overview

A high-level block diagram of the MDISS appears in Figure 2.  The light source, on the left, consists of multiple subsections, including the seeded pump laser, an optical shutter, the OPO, energy leveling, idler beam removal, and beam monitoring components.  On the center right of the figure is the sample holder containing a test cell, which in turn contains the biological sample to be characterized.  Surrounding the sample cell is an energy collection system that simultaneously captures the input energy, and the reverse-scattered, forward-scattered, and unscattered energy.  The control section is depicted at the bottom of the figure.  These four sections comprise the primary hardware elements of the MDISS system.

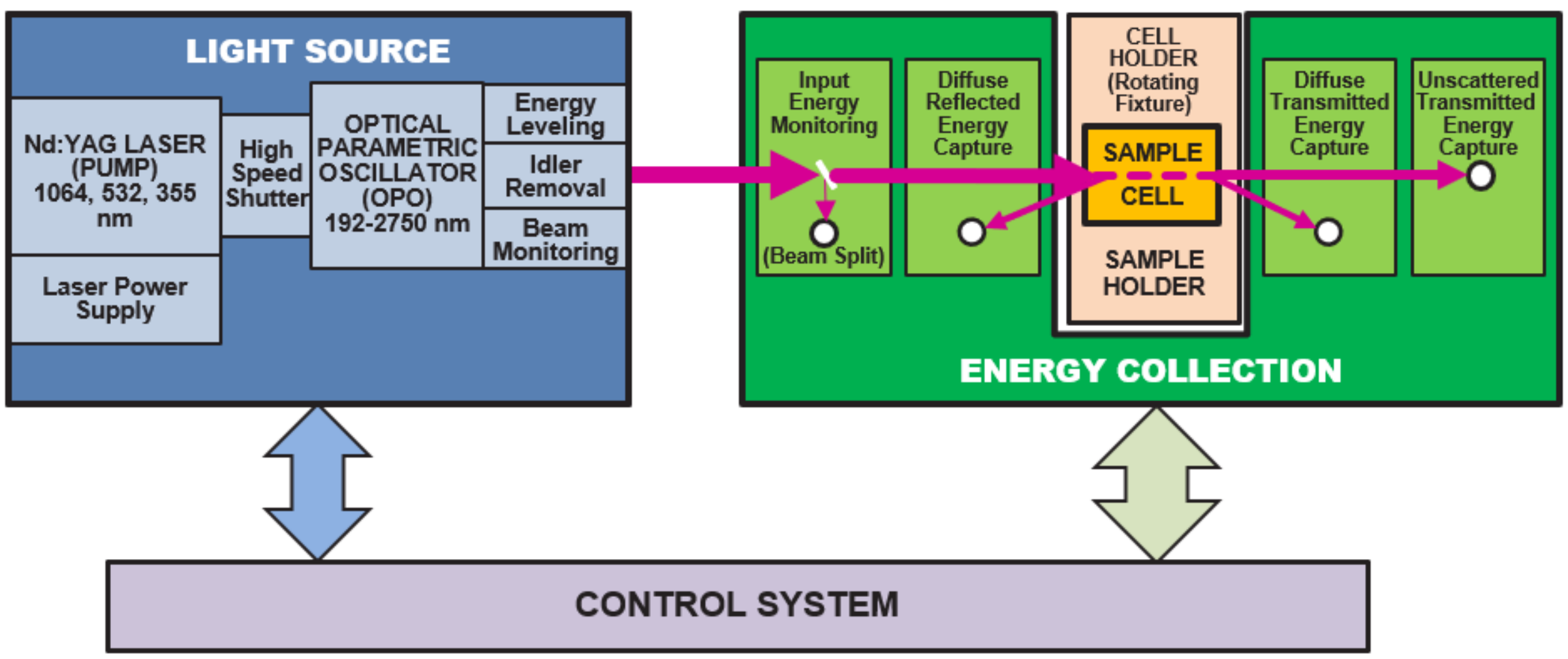

**Figure 2** MDISS block diagram.  The narrow linewidth pump laser coupled to an OPO allows precise optical characterization of biological analytes over a 192–2750 nm wavelength span (45857)

The physical MDISS system appears in Figure 3 implemented on a 4 by 10-foot optical table. The light source is on the right; the sample holder and energy collection system are on the left. The primary building block components have been identified in the figure and will be further described in the following sections.

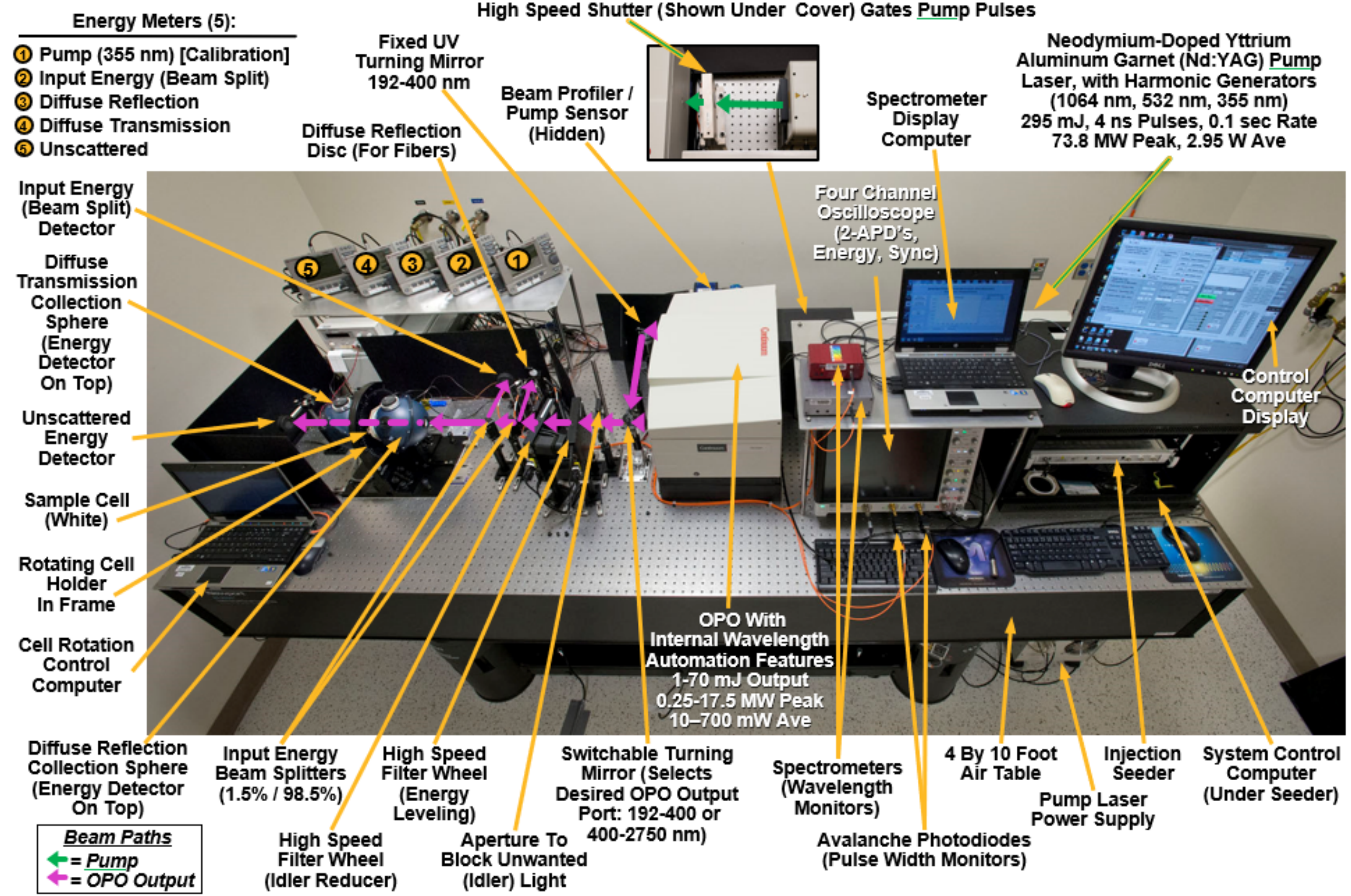

**Figure 3** Physical MDISS system, layout mirrored from Figure 2 (e.g., the light source is on the right in this figure, but on the left in Figure 2), with the individual functional elements labeled. (45428v2)

## 2.2. Details of the functional blocks

Figure 4 provides greater detail of the subsections. The light path begins at the pump laser and traverses the sequence of functional elements. The pump produced ten optical pulses per second that passed through a shutter, allowing the groups of pulses to propagate to the OPO. Using frequency doubling and tripling, three coherent wavelengths were provided to the OPO. A pulse-aligned electrical timing signal from the pump was distributed to the downstream path. With precisely controlled rotations, the OPO crystals converted the fixed pump optical wavelength energy to two wavelengths spanning 192–2750 nm, *i.e.*, the signal and the idler beams where: $1/\lambda\text{pump} = 1/\lambda\text{signal} + 1/\lambda\text{idler}$. For most of the instrument's $\lambda$ range, the so-called signal beam was used; for the rest of the range, the idler beam was used. Physics dictates that the OPO output levels vary with wavelength. The selected OPO output pulse train passed through an energy peak limiting filter and a filter that removed the unwanted part of the spectrum. The resulting output beam was monitored by splitting fixed proportions of the optical energy to energy sensors (nominally 0.5% was tapped at each of 3 for a total of 1.5%). The remaining energy was applied to the sample cell, where the reverse scattered, forward scattered, and unscattered energy were collected simultaneously. Additional monitoring and performance verification functions are depicted in Figure 4 by dashed boxes.

As noted above, the end-to-end system can be subdivided into four functional elements: the light source; the sample holders (multiple types of sample holders to accommodate a variety of samples); the energy collection section (multiple detectors "encapsulating" the sample holders); and the control and data acquisition section (automated hardware and control software). Each will be described separately below.

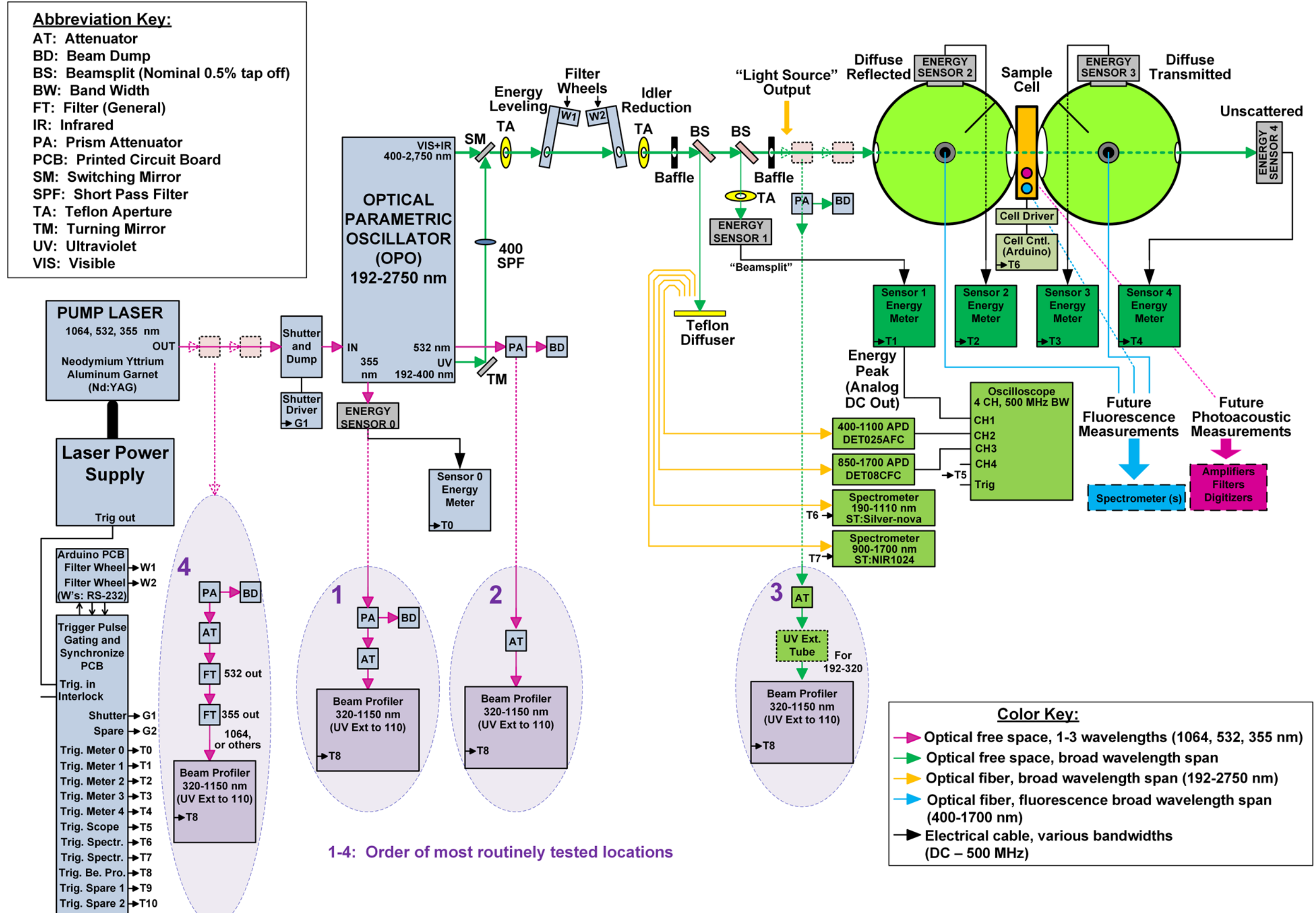

**Figure 4** MDISS detailed block diagram (45882v2)

### *2.2.1 The Light Source: Pump Laser and OPO*

As depicted in Figure 2, Figure 3, and Figure 4 the light source consists of six sections:

1) The Pump Laser is the fundamental source of optical energy. The laser incorporated into our design was a Surelite EX with an SL-200 seeding option, commercially available from Continuum Inc. This light source is a classic flash lamp-based Neodymium-Doped Yttrium Aluminum Garnet (Nd:YAG) pump laser operating at a fundamental wavelength of 1064 nm, which, as noted above, was then frequency-doubled and -tripled, resulting in additional wavelengths of 532 nm and 335 nm. All three wavelengths exited the laser, but only the 355 and 1,064 nm λ light were used by the next downstream component, the OPO. The Surelite generated nominal 355 nm 315 mJ output energy at the nominal 4 ns pulse width (pulse width span is 3-5 ns) is equivalent to a pulse output power of 78.8 MW, and at the 0.1 second pulse repetition rate which is equivalent to an average power of 3.15 watts. Although the Surelite pump is a high-performance source, close examination of the output pulse energy specification demonstrated a variation of ±4% for 99.9% of the pulses (the remaining 0.1% of the pulses exhibited even greater variability). This variation combined with the 6% power drift over an eight-hour duration mandated the need to monitor the output energy on a pulse-by-pulse basis (described later).

2) The Shutter. The pump laser output then passed through a high-speed optical shutter. The pump laser ran continuously to minimize pulse instabilities. Although the pump could be disabled by means of the control software, the high-speed shutter served as a faster and pulse–time–aligned way to gate groups of pulses, e.g., pulse sets of 10, 100, or more. Typically, ten pulses were gated to the OPO for each OPO output wavelength of interest. For example, ten pulses would be applied to produce ten OPO output pulses at 500 nm; the shutter would then close while the OPO optics were changed to produce an output at 505 nm (a 5 nm step); the shutter would then open to allow ten 505 nm pulses to be passed through; and so on. The shutter exhibited a 40 ms open/close time, which was synchronized to operate between laser pulses occurring each 100 ms.

3) The OPO was a commercial Horizon II system from Continuum Lasers, with provision made at the time of purchase for the addition of automation features. The OPO converted the 355 nm light from the pump laser to the desired wavelength for a given measurement through a second-order optical process in a nonlinear optical crystal, the β-phase of barium borate (BBO, BaB2O4). The angle of the BBO with respect to the incident 355 nm input determined the signal and idler wavelengths. The Horizon II control software and associated motor positioning hardware allowed the wavelengths to be changed under user and test program control. Optical components within the OPO housing were used to select either the signal or the idler, depending on the desired output wavelength. With one set of the BBO crystals and the 355 nm input beam, output wavelengths from 400 nm to 2,750 nm were generated and emerged from one exit port of the Horizon II system. To generate output light in the ultraviolet (UV) range (192-400 nm), another BBO crystal set, and a second-order optical process were used by passing the signal beam from the first BBO crystal, along with the 1,064 nm light from the pump laser, through another non-linear crystal; this light was available through a second exit port of the Horizon II. The output energies from the two ports were directed (see Figure 4) into a single beam path that then illuminated the sample under measurement.

Figure 5 presents internal views of the OPO in both the standard manually controlled configuration and with the added automation features, which were incorporated in two specific areas. All the pneumatically operated positioning mechanisms incorporated reed sensors at each of the extreme ends of their travel paths to provide positioning confirmation and feedback. Automation was added external to the OPO, allowing the energy from both output ports to be merged onto a common optical output path directed to the sample. The external element was a switching mirror that was 1) moved out of the optical path for the 400-2750 nm output to be directed downstream to the sample cell, and 2) moved into the path to allow the 192–400 nm output to be directed downstream to the sample cell. This pneumatic control/driver system was electrically integrated with the entire automation system, described later.

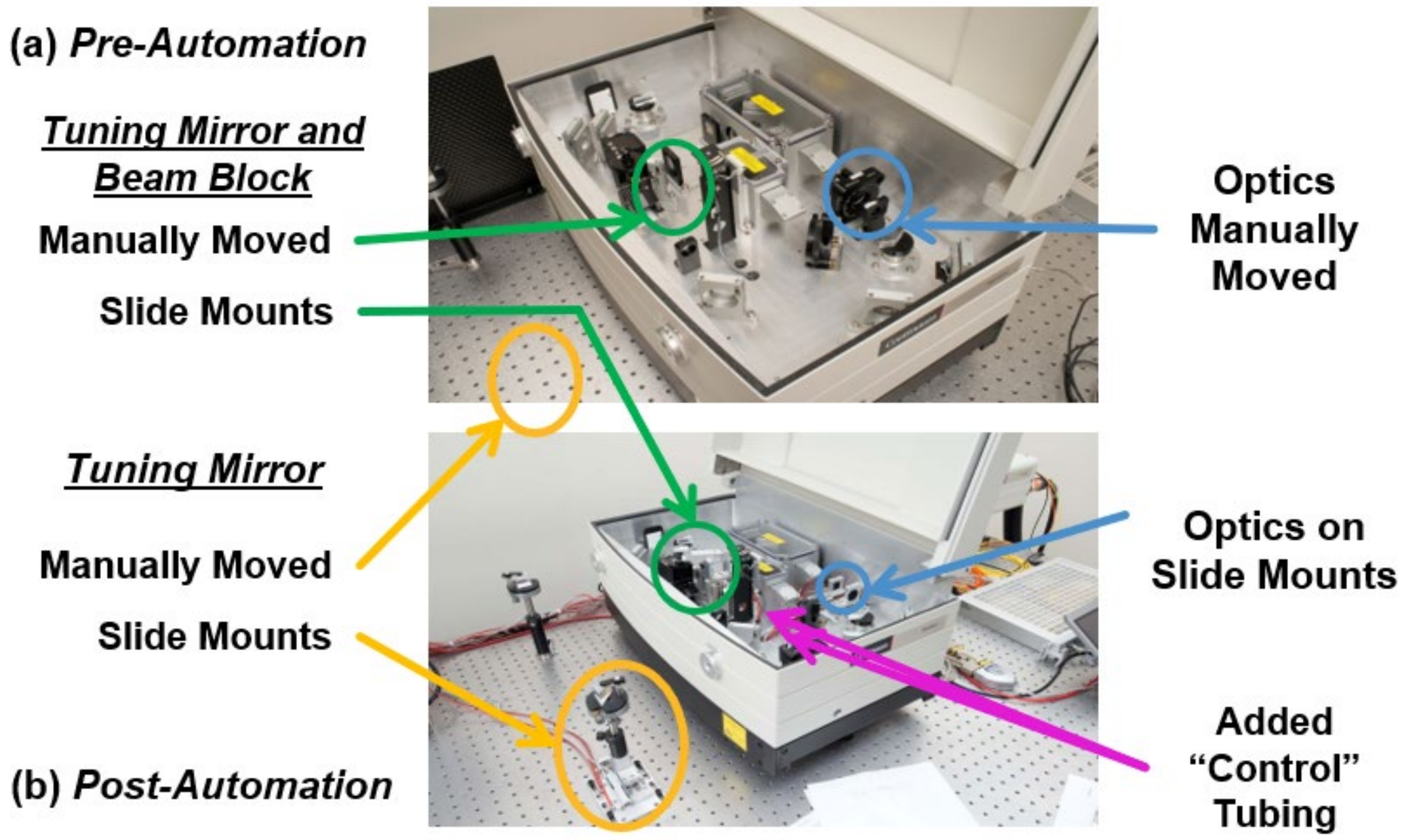

**Figure 5** Internal view comparison of (a) manual versus (b) automated OPO used in MDISS system, operating over a 192–2750 nm wavelength span. Continuum horizon OPO features: 3-5 nm pulse width; less than 100 micro-radian pointing instability; 2–6/3–7 cm seeded/unseeded line width; 4–7 nm beam diameter; less than ±10% energy instability; 192–400/400–2750 nm dual optical ports (45867)

4) Energy leveling. The automated OPO thus provided a 192–2750 nm wavelength span on a single beam path. However, the output energy was not uniform over the wavelength span. A maximum energy limiting approach was implemented by inserting a Finger Lakes Instrumentation model HS0125 ten-position filter wheel in the output path, as illustrated in Figure 6①. The filter wheel was fitted with selected 25 mm diameter optical neutral density filters (i.e., optical attenuators) to limit the output energy to a range of 2-10 mJ, providing more uniform incident energy to the sample. By limiting power to only those levels necessary for the measurements these filters also prevented sample damage. Further, the limited energy levels allowed energy sensor ranges to be fixed enabling the use of optimized energy sensor measurement spans.

5) Idler removal. The energy-leveled beam then passed through a second filter wheel (the same model as the energy leveling wheel) with wavelength-specific bandpass filters ("long pass" and "short pass" in optics terminology) to remove the idler beam (see also Figure 6②).

**Table 1** OPO output specifications, including derived parameters from the pump Laser

| Wavelength span: | 192–2750 nm |
|---:|---|
| Line width: | less than 1 nm up to 2200 nm |
| Pulse rate: | 10 pulses/second (100 ms time interval) |
| Pulse width: | 10 ns |
| Pump source power drift over 8 hours at ±3 degrees C: | ±6% |
| OPO energy stability: | ±10% for 99.0% of pulses |
| Beam diameter: | nominal 8 mm diameter, but varies in size and shape versus wavelength |
| Beam wander: | ±5 mm from center reference (varies with wavelength) |

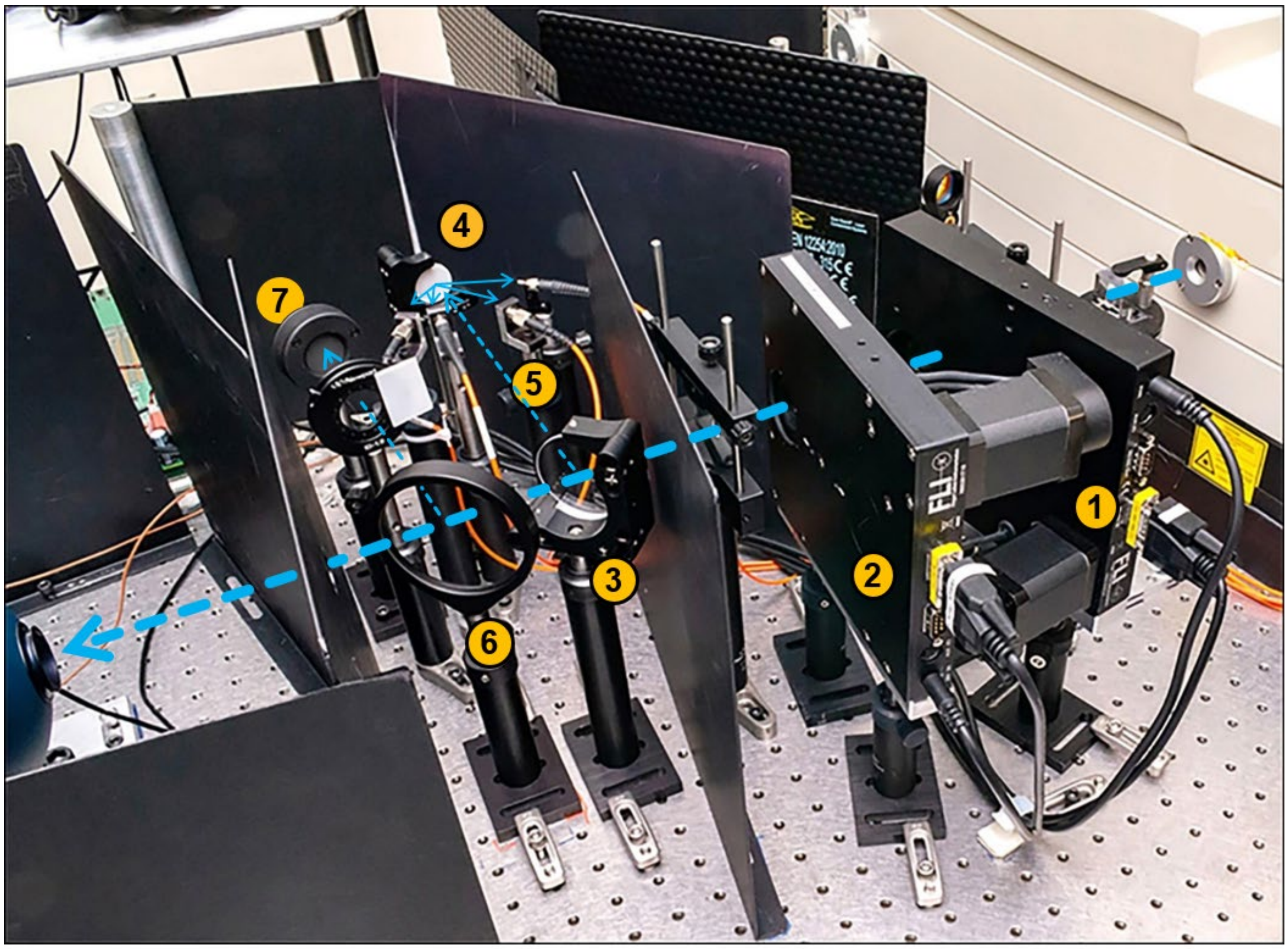

**Figure 6** MDISS OPO output beam monitoring configuration. Two filter wheels (① and ②) were used to control the range of energy and λ delivered. There was a dual beam split on the OPO output path; the first split ③ was to a Teflon diffuser ④ with reflections detected by fibers ⑤ for two spectrometers and two photodiodes, the second split ⑥ was to an energy sensor ⑦ (46547v2)

6) Energy/beam monitoring. As noted in the pump and OPO descriptions, there was an observed variability of up to 10% in the pulse energies at the OPO output, and overall power drift changes of up to 6% over eight-hour time spans. To account for this variability, we monitored the output energy on a pulse-by-pulse basis and used these values to correct the downstream measurements (see a later elucidation of this correction process). However, we also had to measure and correct for other OPO output parameters such as wavelength accuracy and time domain pulse shape conformation. To implement this function, a pair of beam diverters (splitters) were incorporated into the MDISS system (Figure 6). The first splitter ③, a 50 mm diameter 3 mm thick calcium fluoride ($CaF_2$) window, allowed diverted energy to strike a Teflon diffuser ④. Four fiber cables were positioned to capture reflected light from the diffuser to allow the wavelength and the pulse time profile (pulse amplitude versus time) to be monitored ⑤. With reference to Figure 4, this monitoring process employed a pair of spectrometers and avalanche photodiodes (APDs) (each with specific wavelength spans) and an associated oscilloscope. The input energy was monitored with a second beam diverter ⑥ (a 76.2 mm diameter 8.0 mm thick CaF2 window (Newport CaF2-W-76.2x8) in combination with a Coherent ENERGYMAX J-25MB-HE thermoelectric sensor ⑦, which in turn was connected to an energy meter (Coherent FieldMaxII-TOP).

### *2.2.2 The Sample Cell and Cell Holders*

The biological sample to be measured was contained in a thin "pancake"-shaped container referred to as the sample cell; several generations were designed and used for different applications. Figure 7 illustrates the details of the adjustable path length cell type. Figure 7a is an exploded view of a generic fixed-path-length example. Cell bodies capable of different path-lengths may be observed in (b-e), depicting four cells with path lengths of 1, 2, 5, and 10 mm. The cell consisted of a central housing machined from a biocompatible material (poly-ethyl-ethyl-ketone, or PEEK), with a window, O-ring and window locking ring on each side. The cell has an octagonal outer perimeter, with a circular windowed central volume, and two straight through circulation ports through opposing sides (i.e., with no bends or corners) to allow straight-through flow of liquids.

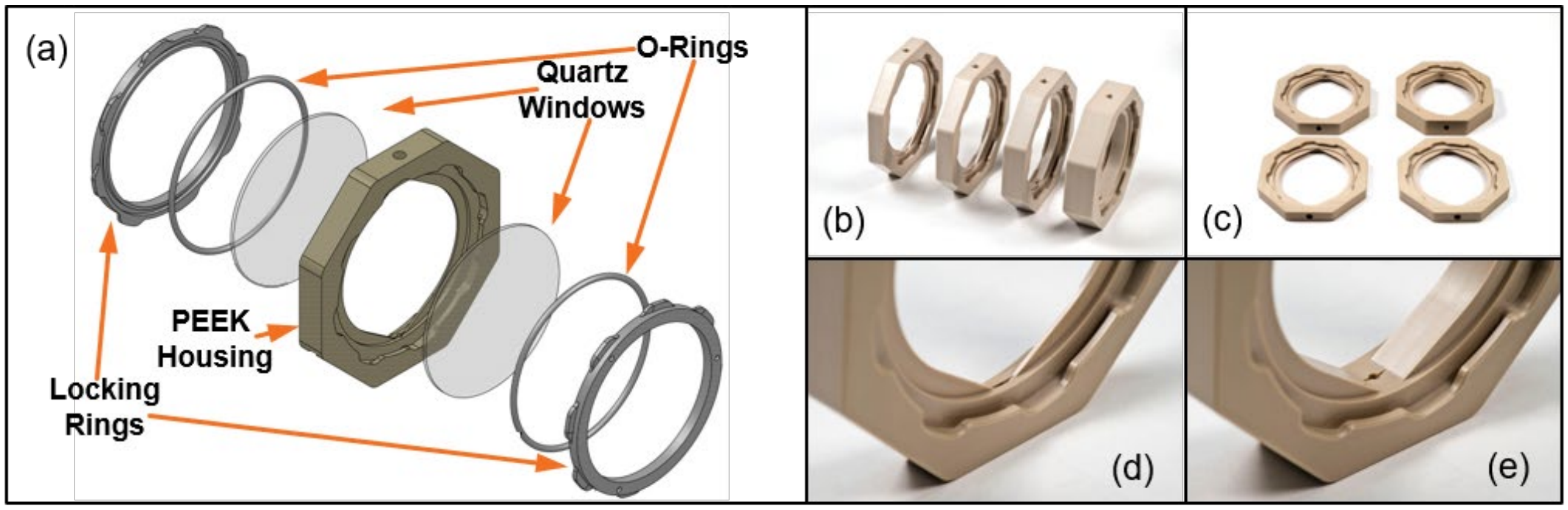

**Figure 7** Custom cell development for MDISS system. (a) a generic exploded view. (b) and (c) illustrate the variable thickness sample cells while (d) is a detail of the 1.1 mm PEEK housing, and (e) is a detail of the housing used for a 100 mm path-length cell. The quartz windows are optically flat. (60221)

Because some of the measurements extended multiple hours, settling and sample stratification was a potential issue. For these tests, the sample cells were mounted in a motor-driven rotating cell holder; details appear in Figure 8. The left portion of the figure (a, b, c, d) presents different views of the cell holder and its constituent components. The right portion of the figure (e, f) illustrates that the sample cell is sandwiched between the scattered-energy collection spheres.

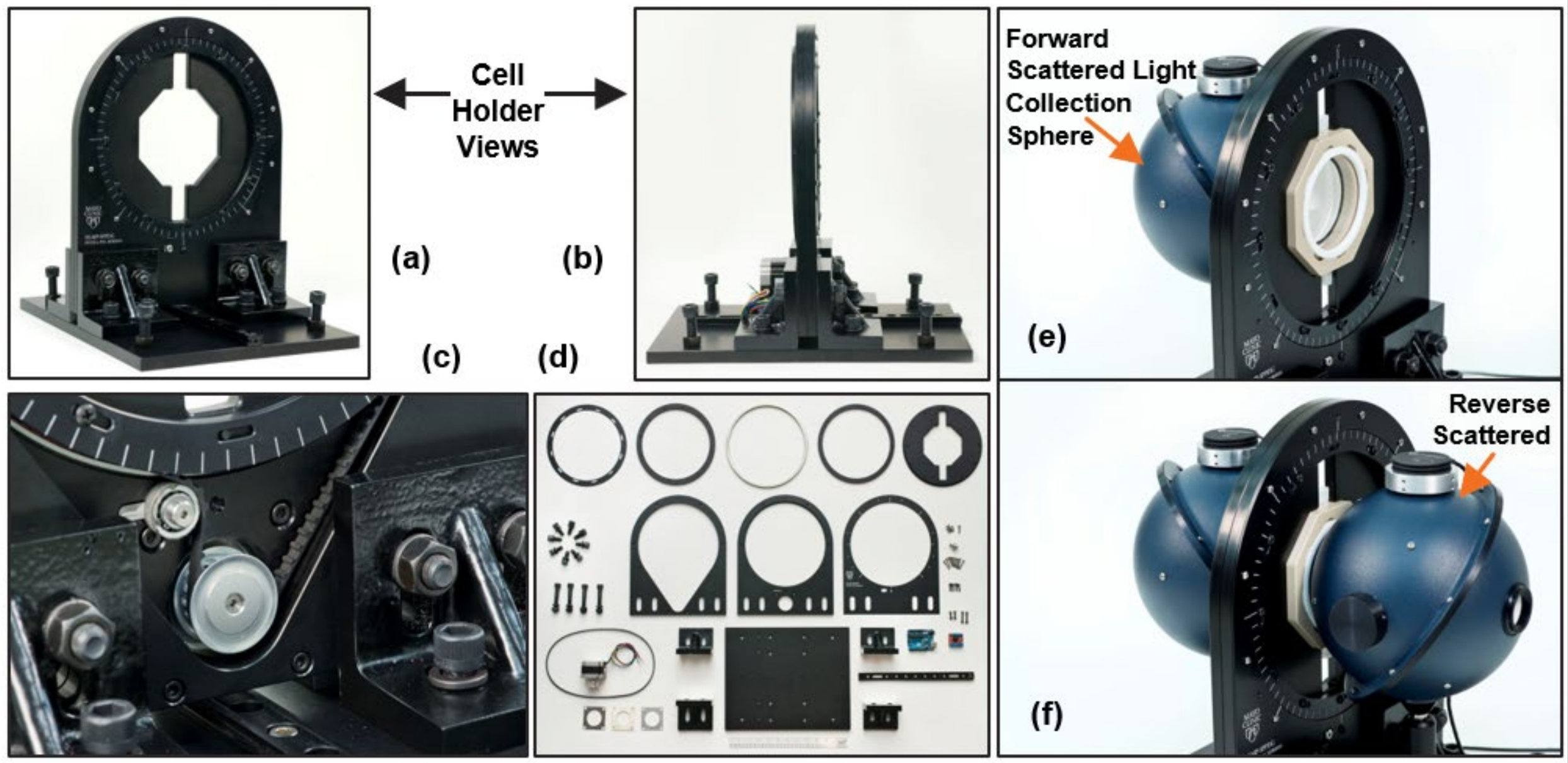

**Figure 8** Custom cell holder development for MDISS system with Forward and Reverse Energy Collection Spheres. (a) Rotation Fixture (b) sideview of same (c) Stepper Motor, Drive Pulley, and Belt (d) Individual Parts (e) view with Forward Scattered Collection Sphere, and (f) with the Reverse Scattered Collection Sphere added. Shown here with 10-mm Path Length Cell visible in (e) (60097)

### *2.2.3 The Energy Collection Section: Multiple Detectors Surrounding the Sample Holders*

The energy collection system is comprised of four primary collection points, with a sensor wavelength span of 190–12,000 nm, thermoelectric sensors with 300 nJ minimum sensitivities, and pulse measurement capabilities up to 1 KHz. These collection points monitored the OPO output energy after energy leveling and idler beam removal filtering. Described below, the measurements include: input energy; diffuse (scattered) reflected energy; diffuse (scattered) transmitted energy; and unscattered (direct transmission) energy.

Figure 9 combines a cross-section diagram and a photo of the combined sample cell, cell holder, and energy collection system, with the sample cell in the center. The sample is depicted in red, the sample cell housing in purple, and the IR quartz windows in green. The sample cell was maintained in place by the cell holder (gray in the schematic, black in the photo). Pressed lightly against the cell windows were two energy collection spheres, which had ports that mated with the window surfaces. The spheres had plastic rings around the port where the cell window contact is made, creating a reasonable though imperfect light seal. Black covers were placed over the spheres/cells to eliminate extraneous light.

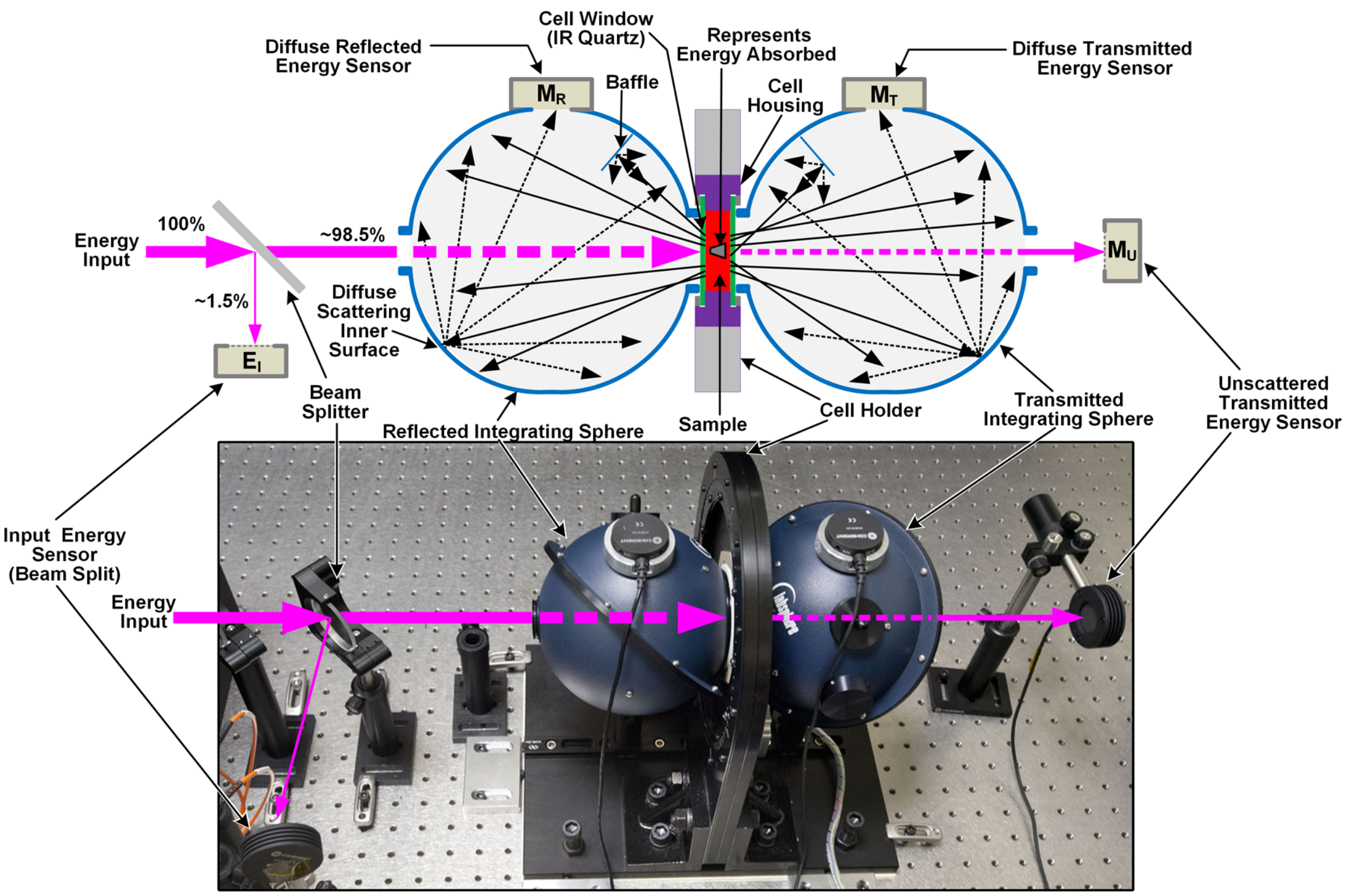

**Figure 9** Depiction of MDISS energy collection subsystem. Beam split sensor monitors input energy from pump laser and OPO; the remaining three energy levels were normalized to the beam split value. (44822)

The primary input beam entered the face of the sample cell, which was sandwiched in the middle of the energy collection system. Beginning at the left portion of the figure, the main input beam is illustrated as passing through the last beam diverter window. Some energy was diverted to the input energy sensor while the remaining main beam passed through and entered a small port of the first sphere, passing through the first cell and then entering the sample cell. Energy could be either reverse scattered, diffuse-scattered reflected, forward-scattered transmitted, absorbed (represented by the small trapezoid depicted inside the sample cell) or passed unscattered to the final sensor. Thus, the sample cells of the multiple types all had to support this "middle of the sandwich" implementation. The baffle inside the integrating sphere prevents straight paths from the input to the detector. The baffle is a diffuse reflecting surface as is the lining of the sphere. A good background in the use and analytics of integrating spheres is [13].

Analog data from the four energy sensors was sent to the energy meters, where the values were digitized and stored, both the raw data values and with a sensor wavelength correction applied on a pulse-by-pulse basis. The integrating sphere energy measured is modified by a factor derived from to the $(\mathrm{Area}_{\mathrm{SPHERE}}/\mathrm{Area}_{\mathrm{SENSOR-OPENING}})(\mathrm{reflection}/(\mathrm{absorption}+\mathrm{reflection}))$. The percentage of light absorbed by the EnergyMax optical sensors was wavelength-dependent and thus the sensors' responses were wavelength dependent. The wavelength compensation in the MDISS programs was performed immediately after acquiring the raw pulse energy data from the meters.

We note that the unscattered transmitted energy measured will contain a small portion of the diffuse-transmitted energy, that which escapes through the direct portal at an angle such that it will fall on the photodetector. This contamination of the measurement is reduced by the cropping ring/spill shown in Figure 10.

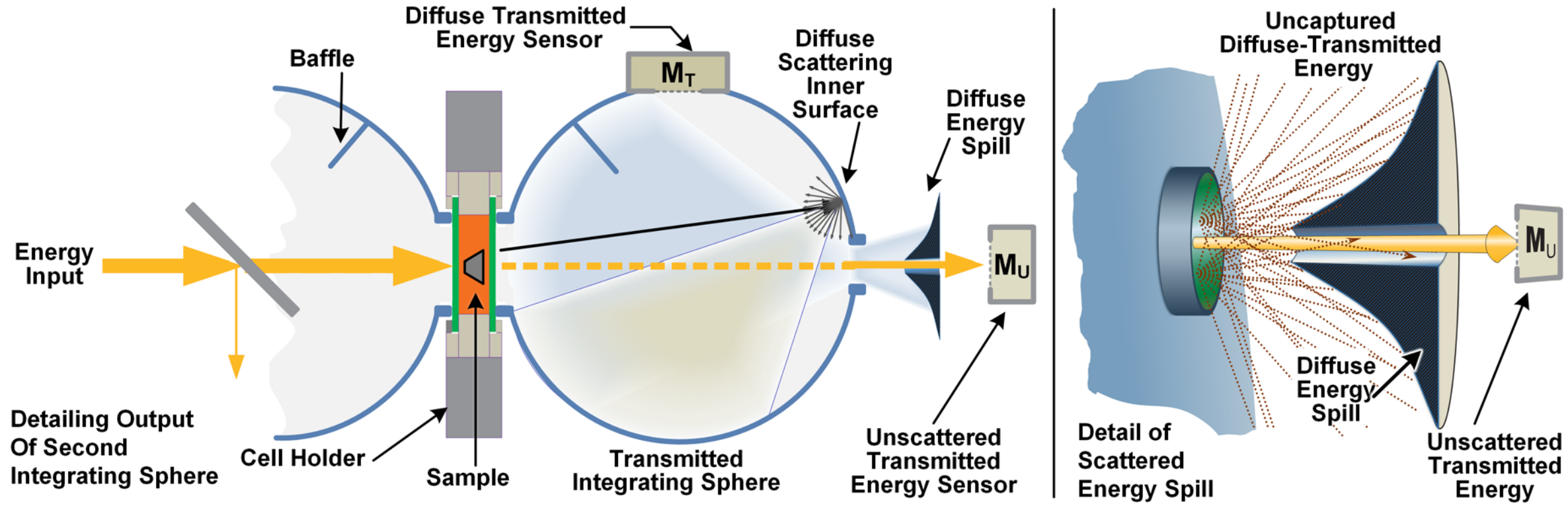

**Figure 10** Using a contoured ring as a shutter to minimize scattered spill onto the unscattered energy sensor. The optical path is from the input energy through … to the unscattered transmitted energy sensor. However, the output port of the second integrating sphere also emits diffuse-transmitted energy. To prevent this diffused energy from impinging on the unscattered transmitted energy sensor, an optical spill reflects the light away from the final sensor. (60220)

#### *2.2.4 The Control and Data Acquisition Section: Automating Hardware and Software*

The MDISS control and data acquisition software was created to satisfy multiple operational requirements, including complete manual control for system performance verification and single wavelength measurements; semi-automatic capability for specific quick tests with limited requirements; and fully autonomous operation for cohort of tests to be run without operator involvement. This high-level control was performed with LabVIEW-based software running with commercial software dedicated for specific functions (*e.g.*, the Continuum Horizon OPO wavelength control software).

The two objectives of the control system were (1) to control the individual elements manually for specific limited tests and (2) to control them in an automated manner to sweep over a range of wavelengths while acquiring energy readings from the meters. We attempted to develop a flexible control system that could be extended to new tests with minimal changes to the software.

Figure 11 includes screen captures for access to the system software. Figure 11a shows the main control window, which controlled the laser power supply/pump and opened the other LabVIEW programs. Secondary window (b) SVP0 provides specific limited tests, while (c) SMP Multi Scan provides control for a cohort of related tests.

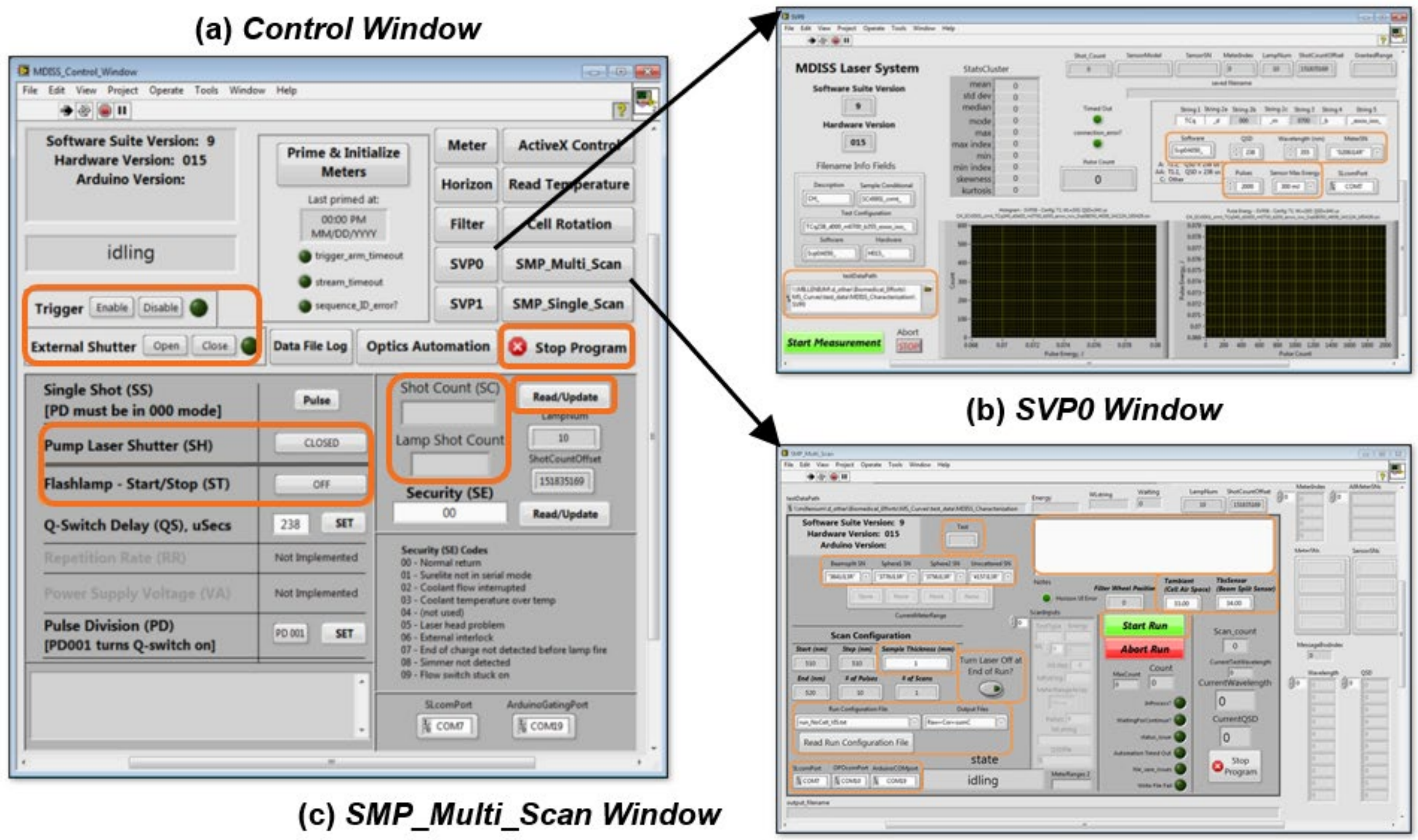

**Figure 11** MDISS primary operating control window (a) with secondary windows (b) SVP0 supporting specific limited tests, and (c) SMP Multi Scan supports staging groups of test runs (60098)

### *2.2.5 MDISS Bioanalyte Measurement Capabilities*

An example of the MDISS characterization capability appears in Figure 12, which is a plot of a subspan (600–1400 nm) of the characterized results for a range of human hemoglobin (Hb) concentrations. Measurement specifics are listed at the bottom of the figure. Eight different hemoglobin concentrations were evaluated, with exact values shown in the figure legend. Note that differences among concentrations were most easily observed in the 850–950 nm span, with essentially no differences observed in the 1250–1350 nm span. This example underscores the importance of an accurate, high sensitivity, broad-wavelength characterization system to determine wavelengths optimally suited for specific characterization activities. The blood samples used to illustrate the measurement capabilities of the MDISS system were collected under full written, informed consent according to Mayo Clinic Institutional Review Board (IRB) study ID 14-001445.

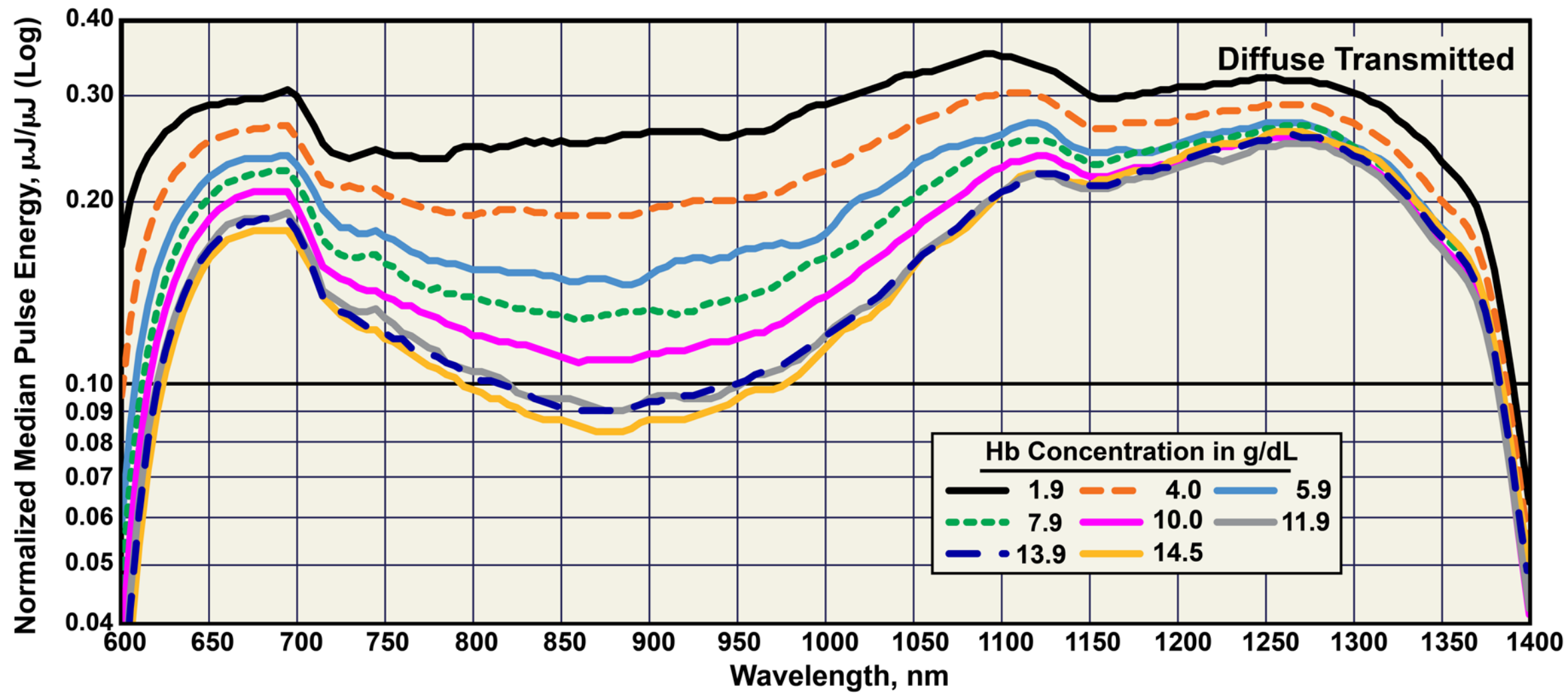

**Figure 12** 100% oxygenated hemoglobin concentration diffuse transmitted measurement. Various whole blood dilutions (using plasma as the dilutant) were recorded with the MDISS in this 600–1400 nm data set (44814b).

## 3. MDISS LESSONS LEARNED, FUTURE RECOMMENDATIONS, AND OPERATIONAL IMPROVEMENTS

The results derived from the MDISS system development and a wide range of bioanalyte test results provided useful data that was incorporated into the design of "downstream" biomedical measurement devices, e.g., as body-worn units. In addition, based on the learning curve derived from the MDISS system engineering and bioanalyte test results, we identified improvements and enhancements that could be applied to next-generation systems. We group these potential improvements into four categories below. We were able to incorporate some of these enhancements; those are indicated thusly ‡.

### 3.1. Light source alternatives and recommendations

New broad wavelength-span light sources have emerged that are continuous wave (CW) implementations rather than those based on pulse/flashlamp combinations. CW lasers provide more stable and predictable optical energy; see, *e.g*., Toptica Photonics. These CW light sources exhibit improved OPO conversion capabilities extending the measurement wavelength range, *e.g.*, diode lasers from Toptica Photonics, using conversion capabilities spanning 190-4000 nm, and fiber lasers spanning 390 to 15,000 nm. These CW light sources provide lower instantaneous power with sustained average power levels. Eliminating the pulse/chirp noise and the power level variations would result in a higher signal-to-noise ratio for the measurements and less damage to the samples.

### 3.2. Energy collection alternatives and recommendations

**a)** Improved energy sensing technologies are now available. The MDISS used pyroelectric sensors; newer alternatives include, *e.g*., Indium Gallium Arsenide (InGaAs) avalanche photodiodes (APD's), Geiger-Mode APDs, and Silicon (Si) dual energy stacked photodiode arrays. As used in, *e.g*., Hamamatsu's S11212-4231/S11299-121 devices [14], these newer photodetectors would provide enhanced sensitivity and improved signal to noise ratios; **b)** photomultiplier tubes could be employed for enhanced sensitivity over selected wavelength spans, *e.g.*, Oriel /Newport #77348, 160-900 nm; #77343, 4000-1100 nm [15]. Silicon Photomultiplier (SiPM) sensors should also be considered *e.g.*, Sense Light (sensL) products; and **c)** multiple energy sensors (such as the above-noted items and pyroelectric sensors) could be instantiated at each sensing location to provide both high sensitivity and broad wavelength span measurement capabilities.

### 3.3. Sample cells: Alternatives and Recommendations

The following extended implementations were or would be beneficial: **a)** Provide the ability to change sample types within a few minutes; **b)** construct a wide range of sample holder types, different sizes, different volumes, and different path lengths, including commercial cuvettes and other sample holders ‡, for a variety of spectrometers; **c)** provide the ability to perform tests at body temperature and at temperatures in the nominal 20-50 C range; and **d)** provide the ability to maintain the sample contents in suspension and/or for the cell contents to be recirculated through the cell ‡.

### 3.4. Operational suggestions:

Following these guidelines: **a)** Maintain an inventory of the highest-likelihood replacement components to support rapid repair (we learned that some critical components were on back order for many months); **b)** develop close collaborative relationships with instrumentation vendors‡, and with a clinical laboratory ‡ to prioritize sample selections, preparation and concentrations with clinical-grade equipment ‡**; c)** perform routine performance tests to monitor the system and obtain a statistically significant performance history; **d)** implement beginning and end of day tests ‡, and/or before-and-after sample tests ‡; support routine system performance tests without requiring physical changes to the system ‡**; e)** use test standards of known analytes (*e.g.*, intralipid-20) to verify system measurements ‡; **f)** acquire a full suite of performance verification instrumentation to characterize the system on a routine basis, including APDs ‡, spectrometers ‡, a beam profiler ‡, and a wave meter ‡**; g)** create system control and automation software in a modular fashion to allow for maximum flexibility as tests proceed ‡; **h)** ensure that operations can be conducted manually, semiautomatically and automatically ‡; **i)** make provisions for tests to be run multiple times with different test configurations in unattended mode ‡, *e.g.*, overnight and over unattended weekends ‡; **j)** provide the ability to perform tests over user-selected wavelength spans, wavelength step sizes, number of energy flashes per wavelength, and number of cycles ‡**; k)** modularize the system configuration into three entities: the light source, the sampling cells, and the energy collection system, enabling the energy collection system and sample cells to be rapidly swapped out and a new or alternate collection system to be installed ‡; **l)** allow for the option to use reduced wavelength span sources and energy collection systems for higher sensitivity tests (*e.g.*, an enhanced test span of 150-300 nm would provide the capability for biological cell wall/cell structure characterization with optimized sensitivity).‡

# 4. MDISS EXAMPLE MEASUREMENTS

The MDISS is able to provide measurments from three fields at each wavelenght, the diffuse-reflected, diffuse-transmitted, and unscattered-transmitted (DR/DT/UT). In this part, a series of stepped measurements made by the MDISS are accompanied by some discussion of the relative observations that may be made.

## 4.1. Sample Preparation

Because the MDISS system required operational verification, the bio-analytes under test were also characterized using clinical-quality commercial instrumentation and techniques. The MDISS measurement results would only be valuable if the detailed characteristics of each sample were known and the MDISS measurements could be matched to known and trusted measurements. For the MDISS optical blood characterization studies, the human blood samples were prepared by specialists from Mayo Clinic's Department of Laboratory Medicine and Pathology (DLMP) using their standard procedures as outlined in [16, cpt. 6], including DLMP's standard practice of dilution of the whole blood down to the opacity that would allow the passage of light. The blood characterization studies were conducted under full written, informed consent according to Mayo Clinic Institutional Review Board (IRB) study ID 14-001445.

The samples were tested by DLMP staff using a clinical quality, calibrated blood gas analyzer (BGA). Specifically, a Radiometer ABL800 Flex BGA was used, as pictured in Figure 15. Immediately after the reference measurement was taken, the test material was inserted into MDISS sample cells and measured. As may be observed in the lower right panel of the figure, the BGA provided multiple blood gas and oximetry measurements. Full BGA-to-MDISS test traceability was implemented, using sample labeling and test cell serial number identification. The sample preparation process and the tests were conducted to minimize the preparation-through-test times to maintain biological stability of the samples. Additionally, a checklist process was implemented to verify the MDISS performance with specific beginning-of-day and end-of-day tests using reference samples, providing a series of calibrations made every day that the MDISS was in operation. The bio-analyte tests were conducted between the calibrations, thereby detecting unintentional and unwanted changes in the system's performance. Lastly, a strict sample cell disassembly, cleaning and cell reassembly process was instituted to ensure measurement consistency. These procedures were created to ensure bio-analyte sample preparation, sample test and MDISS operational uniformity.

## 4.2. Data Series

In the next four sections, several blood analyses are presented. The first analysis characterized blood samples over a range of hemoglobin concentrations; the second measured blood oxygenation using CO2 to deplete the oxygen carried by the hemoglobin (referred to hereafter as a "desaturation" procedure); the third reports the changes with simultaneous hemoglobin concentration changes and blood oxygenation using CO2 as the desaturation gas; and the fourth describes a range of blood oxygen saturation levels using CO as the desaturation gas.

In each section, the MDISS bio-analyte measurement results are depicted in several formats, including different wavelength spans, with some charts presenting all three diffuse-reflected, diffuse-transmitted, and unscattered-transmitted (DR/DT/UT) results, while others present a single result of one of the three measurement parameters. The breadth of information in these charts confirms the utility of a versatile broad-functionality spectrophotometer. The MDISS can efficiently and accurately obtain measurement results for a given bio-analyte sample. In some cases, a broad wavelength span will be described to illustrate both the measurement regions of interest as well as the wavelength spans presenting little useful data. The areas of interest will be magnified to underscore important measurement differences, although the MDISS runs were in fact run over the machine's full-wavelength range in most cases.

***Nota bene***: We make no attestation that the data presented in the charts below are in any sense "ground truth" results. These charts present *experimental data* from an *experimental machine.* These data sets have not been vetted against results in the published literature and should not be taken as such; we are merely illustrating the type of high amplitude and fine-grain wavelength resolution, and in some cases additional parameter generation, that the MDISS can collect. We note that our curves appear similar to those in the literature, with slight differences observed in some sub-waveband regions. Much more work would be needed to reference our data sets against those in the published literature.

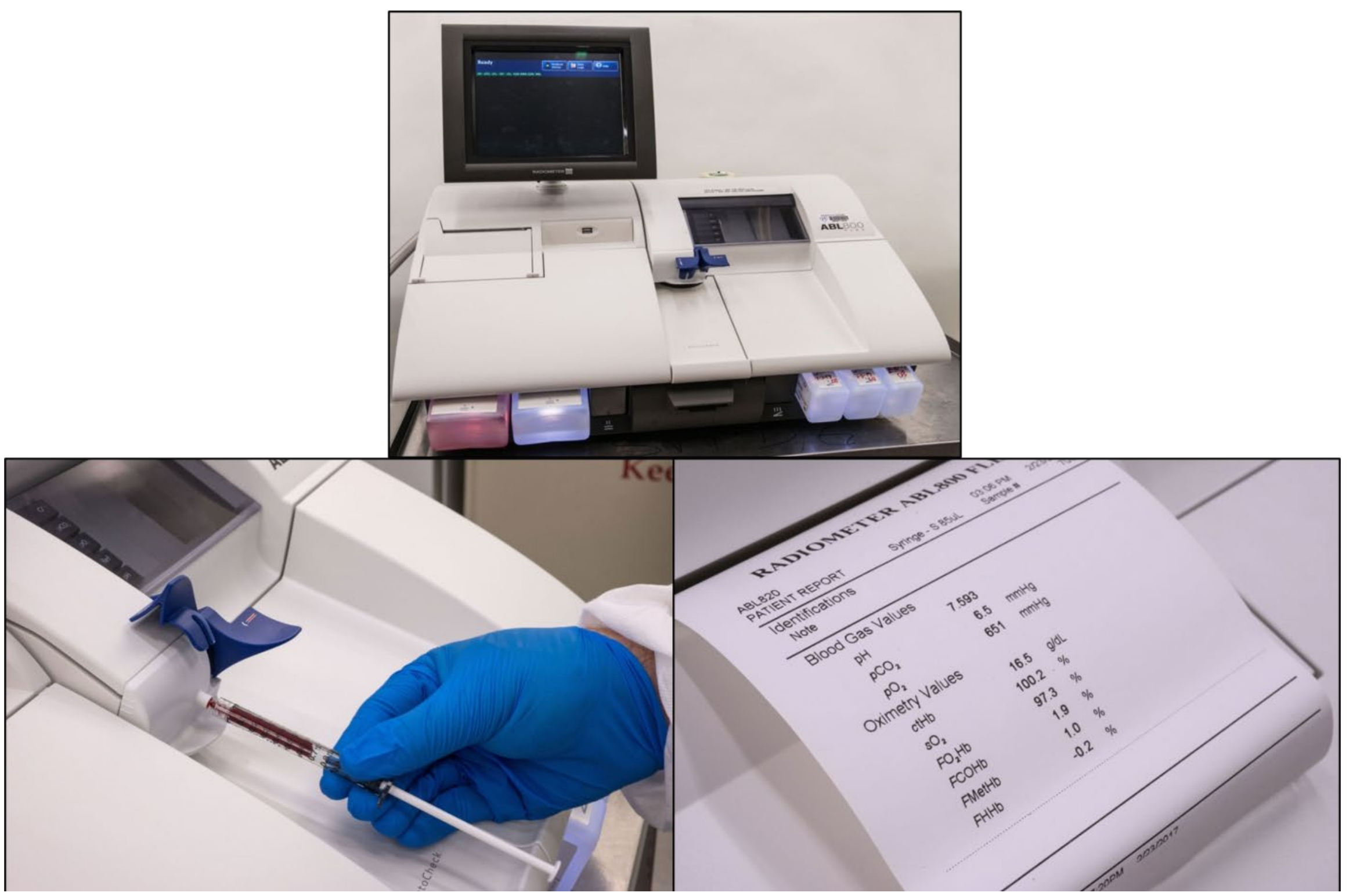

**Figure 13** Radiometer ABL800 Flex blood gas analyzer used in support of MDISS blood characterization (45862b).

### *4.2.1 Blood characterization with different hemoglobin concentrations*

Blood is the most common bio-analyte of interest due to its capability to serve as a broad indicator of overall health status, with well over one hundred unique types of blood tests performed routinely as part of overall reviews of patients' clinical status. The initial MDISS plan was to begin with fundamental tests. The first step was to characterize hemoglobin concentrations at various levels. Figure 14 presents the DR/DT/UT data characterization results for eight different concentrations of hemoglobin over a 400-2750 nm wavelength span. The samples obtained from Mayo's clinical laboratories were diluted with different amounts of plasma, each from the same patients, thereby reducing the effects of other types of sampling variations. Samples with the range of concentration values were prepared, from 14.5 grams per deciliter (g/dL) down to 1.9 g/dL. The sample cell was 63.5 mm in diameter, with a 1 mm optical path length (*i.e.*, a 1 mm path length across the inner walls of the sample cell IR quartz windows). Several features were observed in this data set.

Each of the DR/DT/UT curves exhibit characteristics that illustrate distinguishing features in regions of the wavelength span. The regions vary between the DR, DT, and UT curves. All three DR/DT/UT curves in Figure 14 exhibit the greatest variation, due to concentration differences, in the range of 650-1100 nm. At wavelengths longer than 1100 nm, the DR curves reveal minimal differences. For the DT curves set, there are discernable differences above 1100 nm but with less variation than the UT curves. Areas for the DT with no differences or difficult-to-resolve differences are from ~1900–2050 nm and at 2350 nm and above. For the UT dilution curve set, differences can be observed in the 400-1875 nm regime, and from 2100-2300 nm. In this latter 2100-2300 nm portion, only the lowest concentrations (1.9 and 4.0 g/dL) are visibly different; the remaining six higher concentration values all overlap in this region. In the 400-600 nm portion, all the DR/DT/UT curves exhibit concentration differences, but each has unique characteristics. The DR curves show differences in this 400-600 nm region, with each unique concentration curve just barely resolvable. In the case of the DT curves for this same wavelength span, only the two lowest concentrations (black, dashed orange) are resolvable, and these lowest two concentrations show the most significant differences from one another as well as from the other six concentrations.

Lastly, in Figure 14, considering the UT curves, only the lowest concentration curve (solid black line) is resolvable from the remaining seven concentrations. It appears that even the lower concentrations of the bio-analyte absorb most of the energy; however, at the lowest concentration, a small amount of the DT and UT energies penetrates the bio-analyte in-situ. These hemoglobin curves underscore our original intent for development of the MDISS, summarized below.

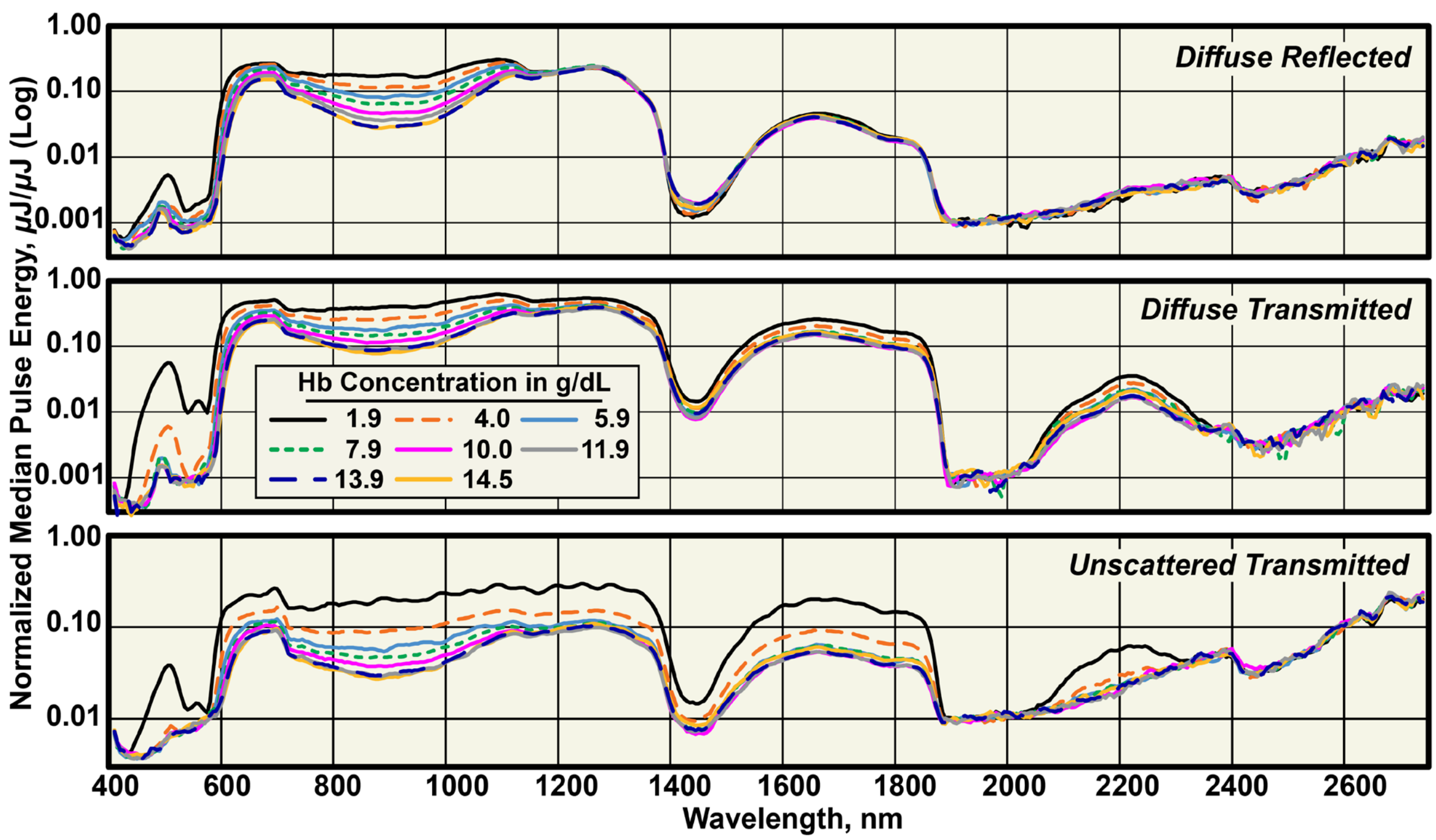

**Figure 14** Various sample concentrations of 100% Fully Oxygenated Hemoglobin measurements (44813).

The MDISS broad wavelength measurement range highlights the portions of the wavelength span suitable and unsuitable for a given analyte measurement. In this case, the wavelengths demonstrating the greatest sensitivity for all three DR/DT/UT curves are the range 800-1000 nm, with the UT curves exhibiting the most sensitivity. However, for the lower concentration values, a wavelength near 500 nm using the DT data may be an appropriate choice, or 1600-1800 nm using the UT curves as another low concentration measurement span. Measurements near 600 nm or from 1900 to 2100 nm or above 2400 nm are not likely to be suitable for the hemoglobin tests for any of the DR/DT/UT curves. We well understand that useful measurements *in vivo* or *in situ* are different from single bio-analyte measurements in vitro. Techniques such as those described in [6] can be used to select wavelengths for measurements of multiple bio-analyte and interfering material components.

Figure 15 displays the DR data from the previous figure, but only in the 600-1400 nm span. When considered relative to the values over 1250-1350, all eight concentration values are resolvable and are ordered top-down from the lowest to the highest concentrations, centered at 900 nm. The two lowest concentration values of 1.9 g/dL (black) and 4.0 g/dL (dash-orange) are resolvable in a small portion spanning 870-890 nm. Based on this figure and visual observation of the curves, a potential wavelength of 880 nm could be suitable for hemoglobin concentration values using the DR data set. Final values would be based on algorithms [6] rather than visual observation of these curves. The ghost curves in the background are the DT curves highlighted in Figure 12. As can be observed, the DR curves provide cleaner separation then the DT.

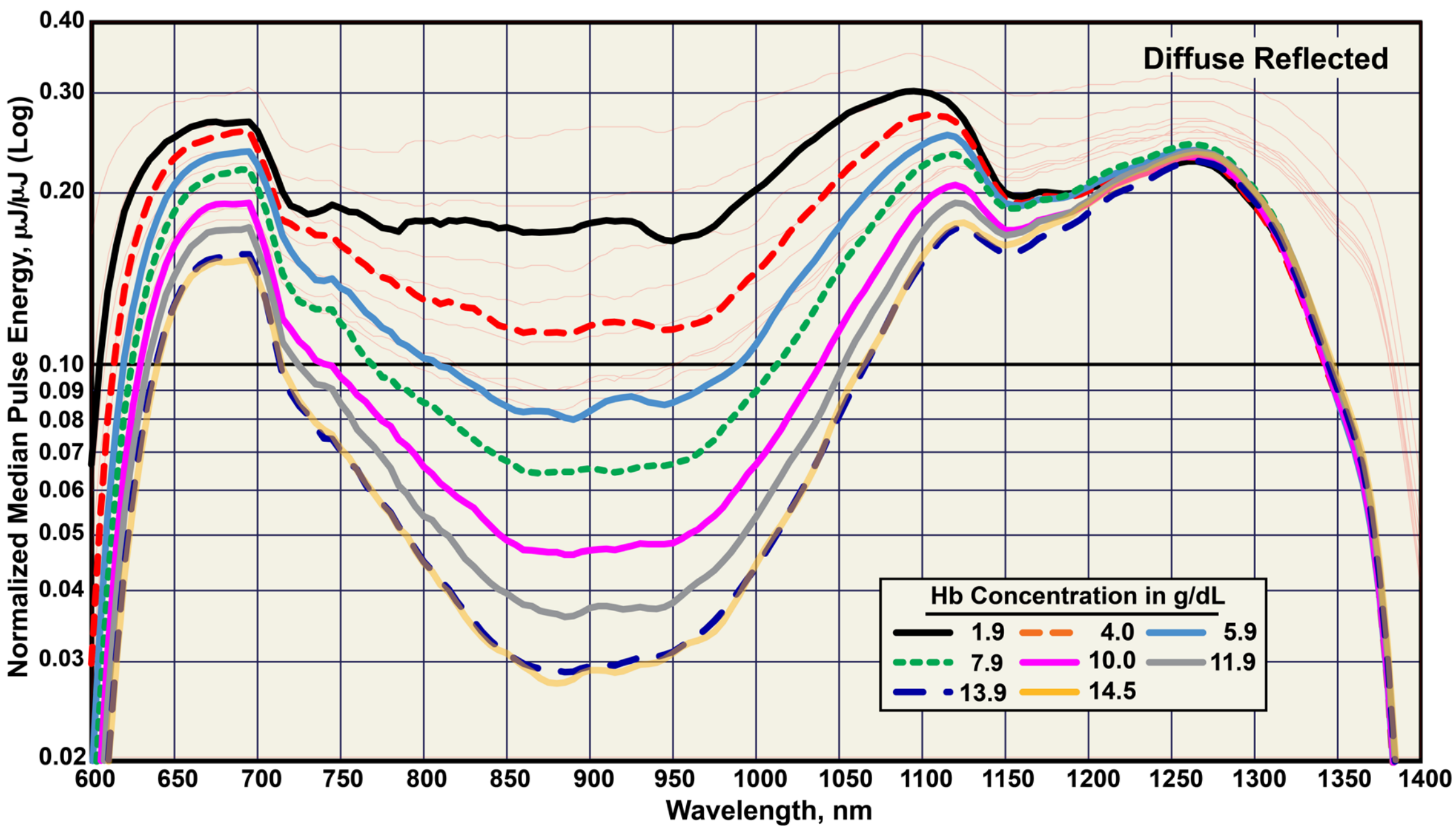

**Figure 15** 100% oxygenated hemoglobin concentration diffuse reflected measurements with the diffuse transmitted measurements shown as ghost curves for comparison (44814a).

### *4.2.2 Blood oxygenation desaturation with Carbon Dioxide*

The next measurement data set comprises a fixed hemoglobin concentration (14.4 g/dL) with changes in the blood oxygenation levels using CO2 as the desaturation gas. Figure 16 displays the DR/DT/UT curves for seven oxygenation values, from 100% down to 3.9%, over a wavelength span of 400-2750 nm. The gas exchange was conducted at 37° C, and the blood tests at 25° C. In this data set, the span showing the most resolvable differences is in the range of 600-1200 nm for all three DR/DT/UT curves. Note that the oxygenation curves cross each other (the dotted black line is the most apparent) at one of the known isosbestic points (*i.e.,* at equal absorption by oxyhemoglobin and deoxyhemoglobin) at a "crossover" wavelength of 805 nm. The other known isosbestic point is at 590 nm, where the curves also cross one another.

An expanded view of a small portion of the data of Figure 16 is presented in Figure 17, which includes only the DR data taken from the previous figure. The isosbestic crossovers may be observed more clearly. The 100% oxygenation curve (dashed dark blue) and the lowest oxygenation (deoxyhemoglobin) curve (black dotted) appear to cross over at 800 nm, near the previously referenced 805 nm isosbestic point. In the 650-700 nm wavelength span, the oxygenation values are ordered top to bottom, with the 100% oxygenation (dashed dark blue) at the top and with the lowest 3.9 % value (dotted black) at the bottom. These order positions invert above the nominal 800 nm crossover point.

Based on this data set and a visual inspection of the curves, good wavelengths for determining the oxygenation values, with CO2 as the desaturation gas, could be 675 nm (below the isosbestic point) and 925 nm (above the isosbestic point). These two wavelengths are different than the previously described 880 nm wavelength suited for hemoglobin concentration changes. In addition, a wavelength at the 805 nm isosbestic point would exhibit no differences due to decreased oxygenation with CO2 as the desaturation gas but would show differences based on variations in overall hemoglobin concentration. These characterization wavelengths are also observed when oxygenation and hemoglobin values are changed simultaneously, as described in the next section.

**Experimental note:** In Figure 16, the UT curve for 85.5% oxygen saturation illustrates a data anomaly in the wavelength range 2520–2540. In this range, significantly more energy was measured than in the surrounding ranges or with other concentrations. The MDISS is an experimental implementation; while the system is under development, there will be occasional data glitches.

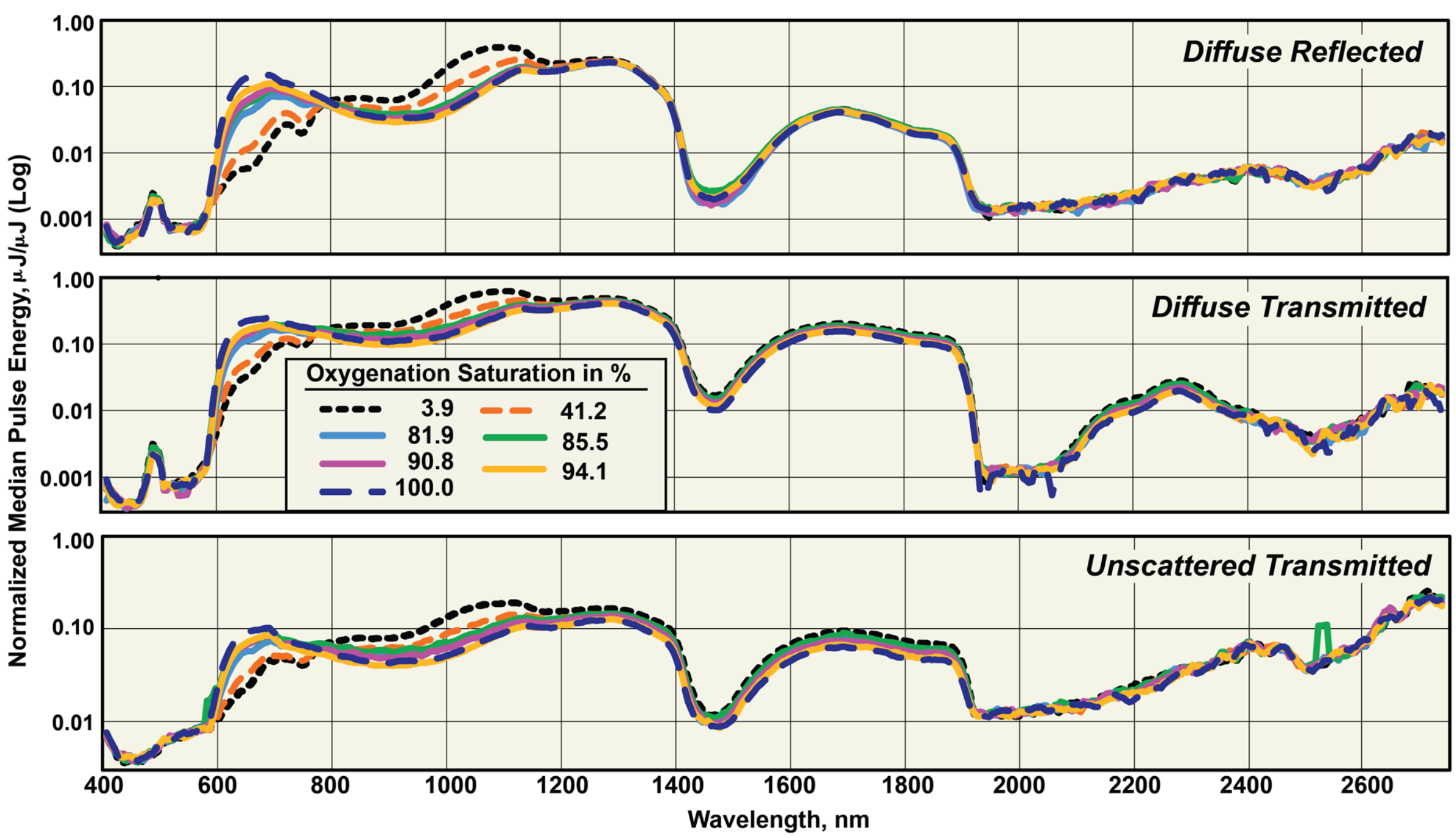

**Figure 16** Carbon dioxide-mediated desaturation of oxygenated hemoglobin measurements in whole blood (44816v2).

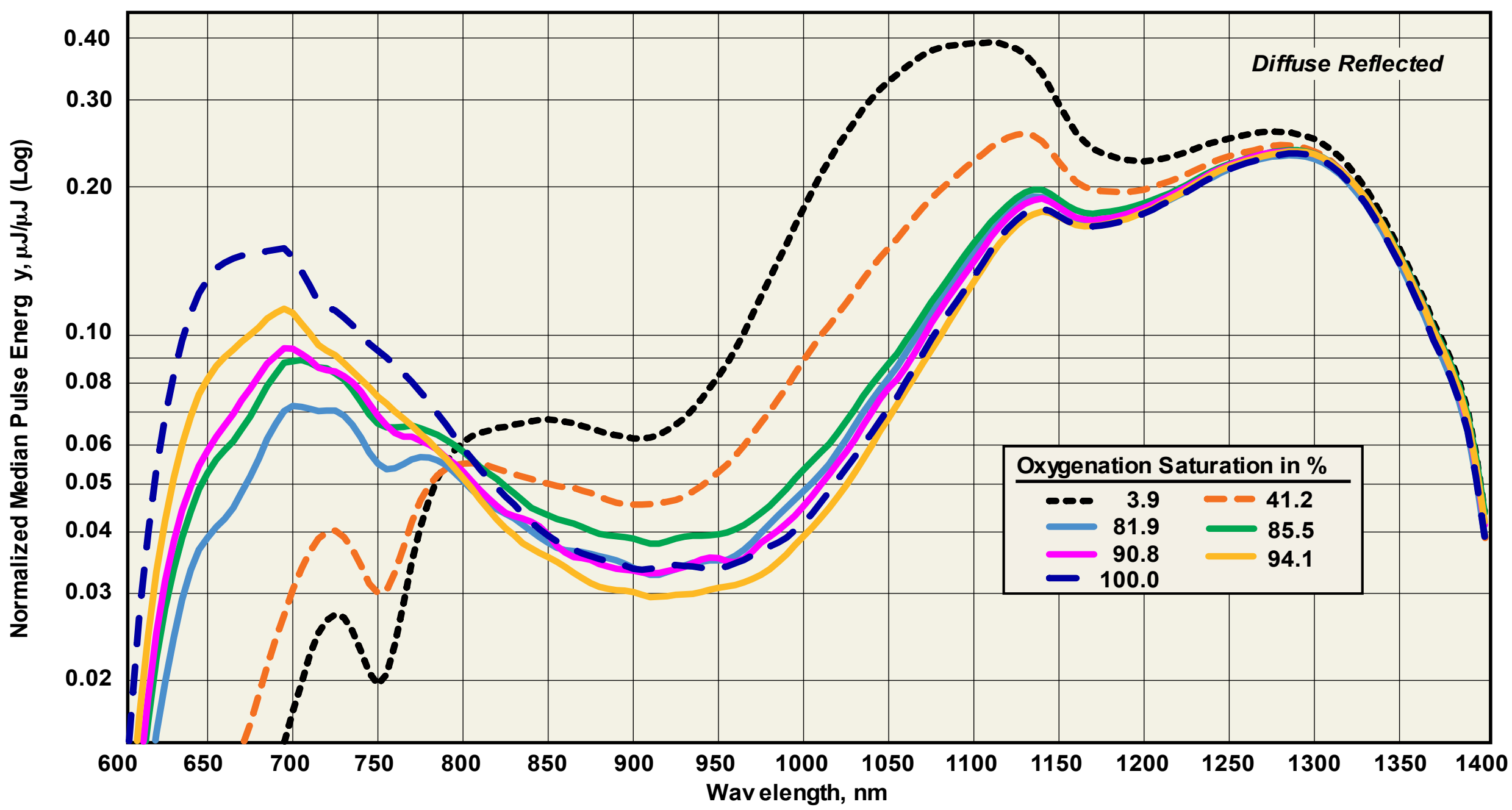

**Figure 17** Carbon dioxide-mediated desaturation of oxygenated hemoglobin diffuse reflection measurement in whole blood (44817).

### *4.2.3 Hemoglobin Concentration Changes and Blood Oxygen Desaturation with Carbon Dioxide*

Here, the results of a change in two independent variables are described: in hemoglobin concentration and in oxygen saturation; see Figure 18. These panels document several macro-effects of variations in these two independent variables. The three left panels are the results with changing oxygenation levels with a fixed hemoglobin concentration of 12 g/dL. The three right panels depict similar changes in oxygenation, but with a fixed hemoglobin concentration value of 6 g/dL. Note that the oxygenation values for the left panels do not exactly match oxygenation values for the rightmost panels. Although these oxygenation levels are close to one another, it was difficult to obtain an exact oxygenation match for the two blood dilutions. The blood was taken from the same IRB registered consented sample source.

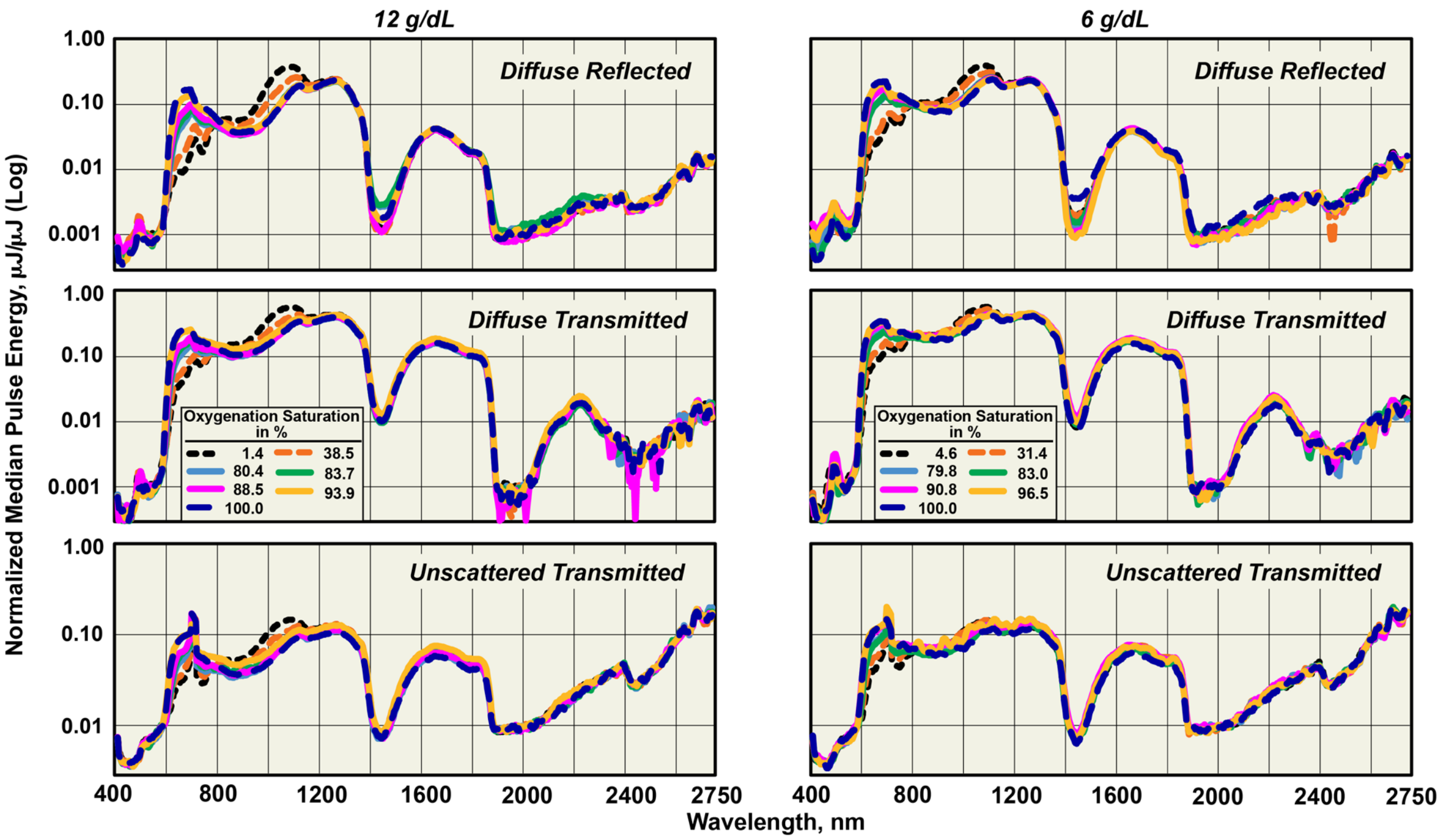

**Figure 18** Carbon dioxide-mediated desaturation of oxygenated hemoglobin in whole blood at 12 g/dl and at 6 g/dl concentrations DR/DT/UT measurements (44825v2).

Two primary results may be observed from this data set. We emphasize that the following are *observations*, without intent to purport deeper meaning to them as of this writing. First, the blood oxygenation isosbestic points, with CO2 as the desaturation gas, at nominally 800 nm and 600 nm do not change; this is as expected. A change in hemoglobin concentration should not affect these isosbestic points, as illustrated in the data sets. Second, the hemoglobin concentration changes do cause small shifts in the suite of curves, which are most apparent in the 800-1000 nm segment in each curve set. For example, in the top pair (DR) of curves sets, the left side curves (12 g/dL) are below the 0.1 pulse energy line at the 800 nm line; on the right side the curves (6 g/dL) are at or above the 0.1 pulse energy line, also at 800 nm. Note that in Figure 16, the lower concentration curves were shifted upward in this 800 nm region, which matches the trends appearing in Figure 19: the 6 g/dL curves are higher than the 12 g/dl curves at 800 nm. For the center panels (DT data), the 12 g/dL curves on the left side are lower (at or just above the 0.1 energy line) at 800 nm than the right side (6 g/dL), where the curves are well above the 0.1 energy line, also near 800 nm. A similar observation may be made for the bottom panels (UT); the left side curves are shifted downward with respect to the right-side curves in the 800-1000 nm wavelength span. From a visual observation of these curves, 800 nm (the CO2-desaturated oxygenation isosbestic point) and 880 nm could be used for hemoglobin concentration changes, and 675 nm (below isosbestic point) and 925 nm (above isosbestic point) could be used for the CO2-desaturated oxygenation characterization. These observations would benefit from future study and refinement.

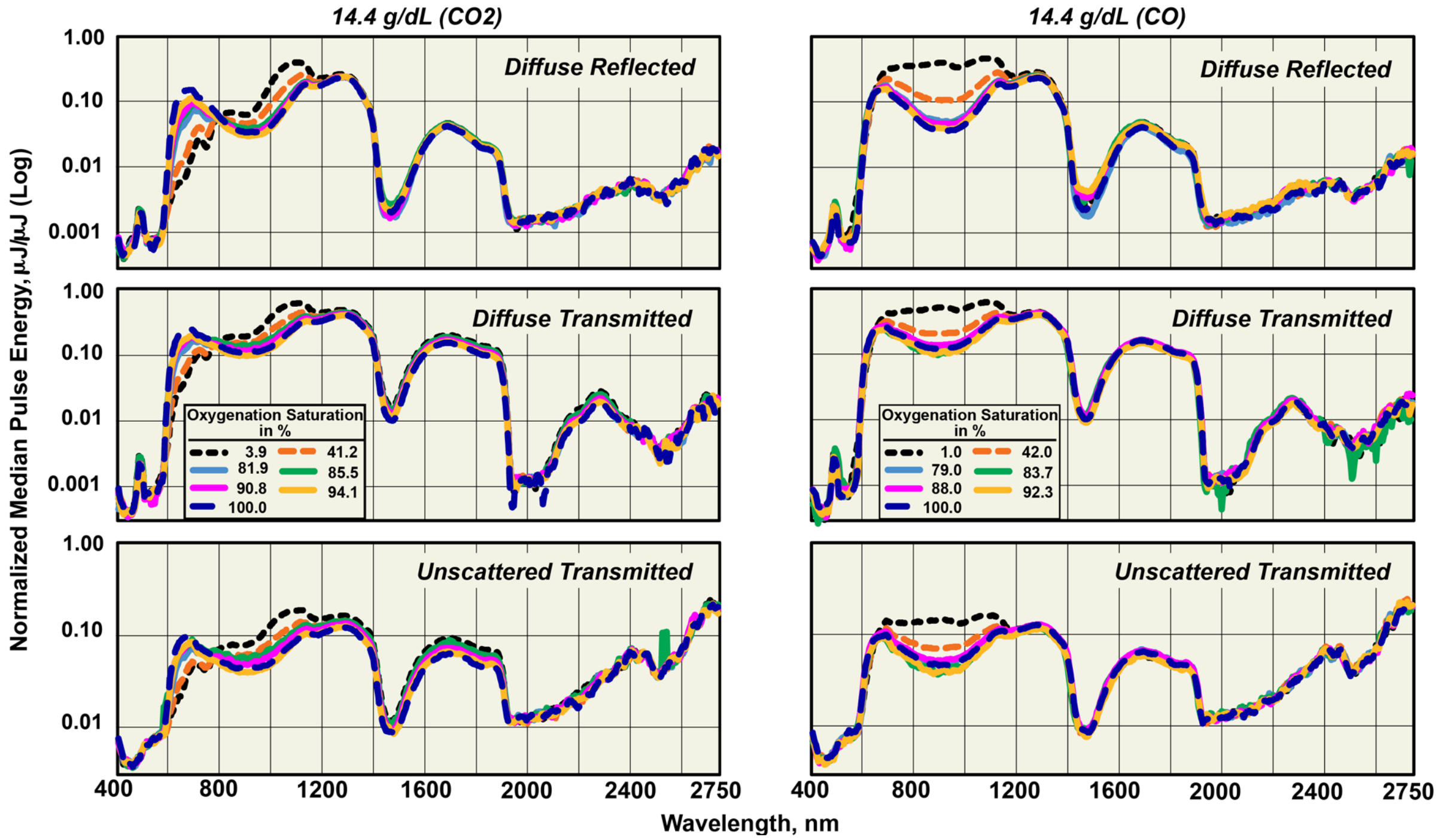

**Figure 19** Comparison of carbon dioxide-mediated and carbon monoxide-mediated desaturation of oxygenated hemoglobin in whole blood at a concentration of 14.4 g/dL (44826).

#### *4.2.4 Blood Oxygenation with Carbon Dioxide and Carbon Monoxide As the Desaturation Gasses*

Another high-level data set is presented in Figure 19. In this case the hemoglobin concentration was fixed at 14 .4 g/dL, but the seven oxygen saturation levels were achieved with two desaturation gases. On the left side are the curves using carbon dioxide (this is the same data previously presented in Figure 17). The curves in the righthand panels were generated using carbon monoxide as the desaturation gas. We present two primary observations. First, for the CO-mediated curves, the changes in oxygenation are much more pronounced; for example, the changes in amplitude are much larger in the 700-1200 nm wavelength band in the right panels compared to the left panels. Note however that the 100% oxygenation curves, *i.e.*, the large-dash black line at the lowermost edge of the curves in the 800-1200 nm region (not the short-dash black line at the uppermost edge of the curves) for both data sets are the same, as would be expected, since these large-dash curves represent no desaturation in both cases. Also note that the right-side CO-mediated blood reveals only a single isosbestic point at 600 nm, whereas the CO2-mediated curves display two isosbestic points at 600 and 800 nm, as previously described. A higher resolution view of a subset of this data is described next.

A comparison of the DR curves for blood exposed to both CO2 and CO is presented in Figure 20. The CO2 mediated blood curves in the top panel illustrate the expected isosbestic point near 800 nm. For the CO-mediated curves in the bottom panel an isosbestic point is present at 650 nm, significantly different than the CO2 isosbestic points at nominally 800 nm and 600 nm. With respect to the high-level MDISS goal of providing design information for small body worn biomedical monitors, these two curves provide the data that would allow "optimal" wavelength selection for a body worn device for CO monitoring (*e.g.*, for body-worn units intended for firefighters). For CO-contaminated blood, by visual observation of the curves we note that 625 nm (just below the CO isosbestic point), and 950 nm will exhibit the greatest sensitivities. Note that there are no distinguishable differences in the vertical axis measurements at the 650 nm CO-mediated isosbestic point, and minimal differences above 1300 nm. In the best case, the optimum wavelength selections for the design of body-worn units should be algorithmically determined, which was not performed on these data sets.

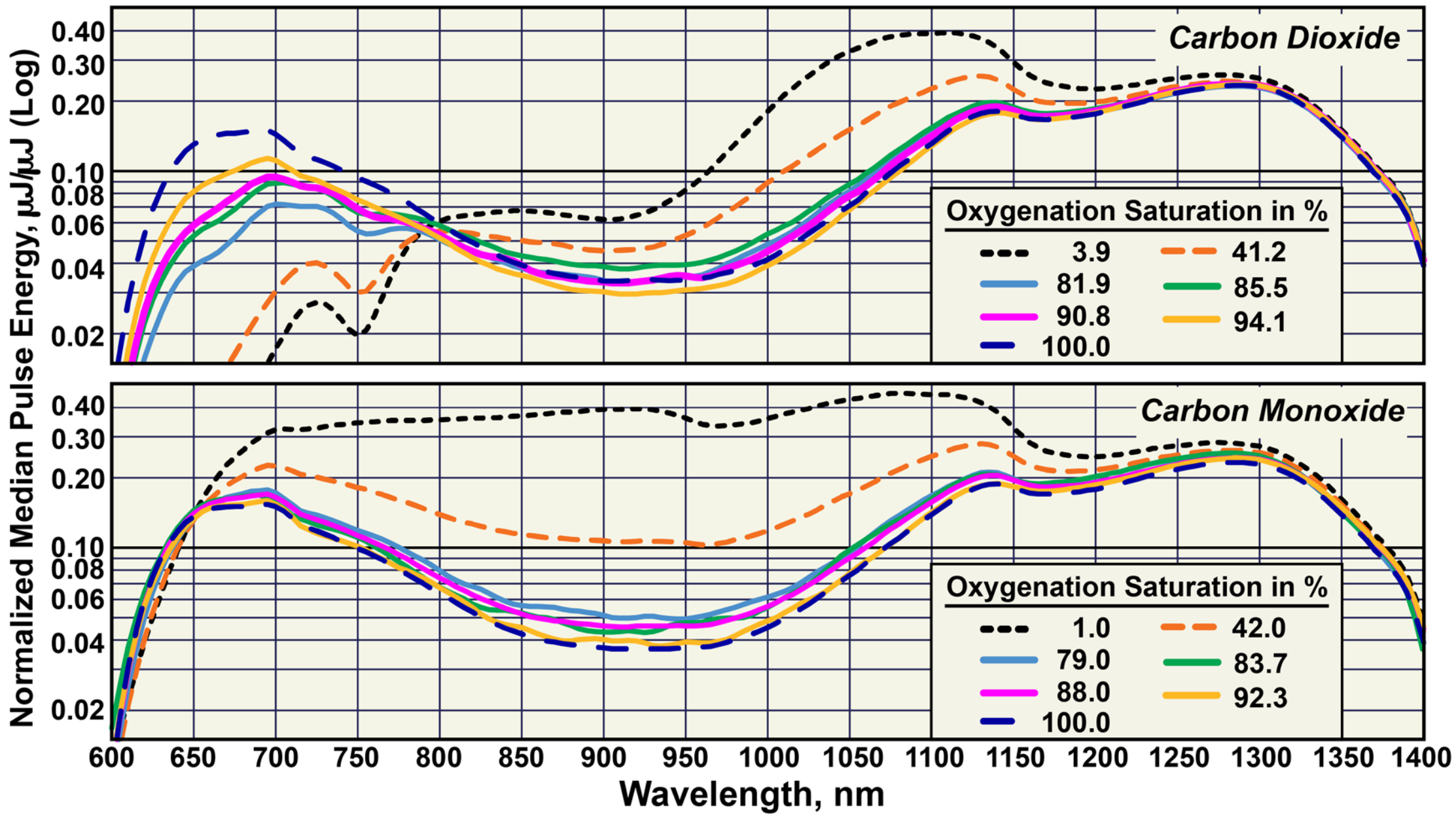

**Figure 20** Comparison of carbon dioxide-mediated and carbon monoxide-mediated desaturation of oxygenated hemoglobin diffuse reflection measurement in whole blood at a concentration of 14.4 g/dL. Note the difference in isosbestic values, approximately 800 nm for CO2 desaturation and 650 nm for CO desaturation (44827).

# 5. CONCLUSIONS

## 5.1. Wavelength Selection Example Summary for Three Blood Analytes

The above set of curves have presented variations in blood gas behavior based on three different variables: 1) hemoglobin concentration; 2) blood oxygenation with CO2 desaturation; and 3) blood oxygenation with CO desaturation. Wavelength selections based on visual observation of the curves, rather than an algorithmic analysis, could be the following: For hemoglobin concentration, ~880 nm could have the largest sensitivity, whereas ~1250 nm would show only slight distinction. For CO2-based desaturation of oxygen levels, ~675 nm and ~925 nm appear to have the largest sensitivity, with no discernable change at 800 nm. For CO-mediated desaturation, 625 and 950 nm would have the largest sensitivity, with no discernable change at 650 nm. We must emphasize that wavelength selection for the light sources and photodetectors for body-worn units designed in the future should be based on algorithmic analysis [6] rather than on these simple, visual observations. However, the more accurate, detailed, and repeatable the data are for these algorithms to work upon, the more accurate, detailed, and repeatable will be the results. It is this task for which the MDISS was designed.

## 5.2. Clinical Perspective

From a clinical perspective, the results of the non-invasive hemoglobin measurements suggest that accurate transcutaneous hemoglobin measurements are likely feasible. The differences in curves between the different hemoglobin concentrations are substantial. Moreover, the differences are larger the lower the hemoglobin concentration is, a highly desirable feature as this type of result might allow accurate and rapid assessment and monitoring of blood loss.

A wearable device or simple sensor that can be used for point-of-care or point-of-patient can be envisaged, and non-invasive hemoglobin monitoring would also be possible, like what is currently done with light-based finger oxygen sensors. As is illustrated in the various O2, CO2 and CO mixing experiments, the capabilities of finger O2 sensors might also be improved, allowing for example immediate, noninvasive diagnosis of carbon monoxide poisoning as well as simultaneous assessment of its severity. At the same time hemoglobin concentrations could be obtained.

Taken in their entirety these studies demonstrate a significant potential to extend these optical experiments to additional analytes and ultimately to a variety of versatile non-invasive sensors.

# 6. Summary

The engineering development of a high-performance spectrophotometer has been described. We view this machine as a multi-generational conceptual descendent of both the Beckman DU spectrophotometer first introduced in 1941 [17] [18] and a successor instrument developed by Hewlett Packard described in [19]. The Mayo unit was developed based on known clinical analyte measurement needs, since no commercial off-the-shelf solutions could be identified. The spectrophotometer implemented a combination of unique features including broad wavelength measurement span, high wavelength accuracy and precision, high sensitivity, rapid, high throughput automated measurement capability, simultaneous acquisition of forward scattered, reverse scattered and unscattered energy, and the flexibility to accommodate a wide variety of sample holder types and volumes.

Multiple bio-analyte measurement results obtained with the Mayo Double-Integrating Sphere Spectrophotometer (MDISS) have been presented, with initial results in several measurement parameters. These results have included hemoglobin concentration changes and blood oxygen saturation changes with two different desaturation gases (CO and CO2). The MDISS system has demonstrated its ability to provide not only these results but also considerable information required to select the wavelengths to detect analytes using standalone clinical-grade body-worn monitors.

### *Disclosures*

The authors declare no conflicts of interest. Although various patents cover various aspects of this work, none of these patents are licensed for gain or profit.

### *Acknowledgements*

We wish to acknowledge the multi-year contributions of the following individuals to this project: Charles Burfield, Anthony Ebert, Theresa Funk, Nicholas Klitzke, Steven Polzer, Jason Prairie, and Steven Schuster; and Drs. Franklyn Cockerill, Kendall Cradic, Graham Cameron, E. Rolland Dickson, Ravinder Singh, and Nathan Harff.

### *References*

[1] Gilbert, BK, CR Haider, DJ Schwab, and GS Delp, "Measuring Physical and Electrical Parameters in Free-Living Subjects: Motivating an Instrument to Characterize Analytes of Clinical Importance in Blood Samples," SPPDG, Mayo Clinic, Rochester, MN, Mayo-SPPDG-R22-15, Dec 2022. Arxiv:2301.00938 doi:10.48550/arXiv.2301.00938

[2] Barker, Steven J and Kevin K Tremper: "Pulse Oximetry: Applications and Limitations," *Advances in Oxygen Monitoring, International Anesthesiology Clinics*, 25(3):155-175, Fall 1987.

[3] Tremper, Kevin K and Steven J Barker: "Pulse Oximetry," *Anesthesiology*, 70(1):98-108, Jan 1989. doi:10.1097/00000542-198901000-00019

[4] Lamego, M, M Diab, and A Al-Ali, "*Physiological Parameter Confidence Measure*." US 7,957,780, issued 7 Jun 2011.

[5] Haider, CR, JA Rose, GS Delp, and BK Gilbert, "Spectrometric Systems and Mechanical Methods for Improved Focus Localization of Time and Space-Varying Measurements." US PTO 9,714,900, issued 11 Sep 2018.

[6] Haider, CR, GS Delp, and BK Gilbert, "Method and Apparatus for Selecting Wavelengths for Optical Measurements of a Property of a Molecular Analyte." US 9,714,900, issued 25 Jul 2017.

[7] Schwab, DJ, CR Haider, CL Felton, ES Daniel, OH Kantarci, and BK Gilbert: "A Measurement-Quality Body-Worn Physiological Monitor for Use in Harsh Environments," *American Journal of Biomedical Engineering*, 4(4):88-100, Apr 2014. doi:10.5923.j.ajbe.20140404.03

[8] Rehman, A-u, I Ahmad, and SA Qureshi: "Biomedical Applications of Integrating Sphere: A Review," *Photodiagnosis and Photodynamic Therapy*, 31:101712, 1 Sep 2020. doi:10.1016/j.pdpdt.2020.101712

[9] Zamora-Rojas, E, B Aernouts, A Garrido-Varo, D Pérez-Marín, JE Guerrero-Ginel, and W Saeys: "Double Integrating Sphere Measurements for Estimating Optical Properties of Pig Subcutaneous Adipose Tissue," *Innovative Food Science & Emerging Technologies*, 19:218-226, 1 Jul 2013. doi:10.1016/j.ifset.2013.04.015

[10] Andre, R, F Moritz, D Klaus, H Andreas, and JM Gerhard: "Optical Properties of Circulating Human Blood in the Wavelength Range 400-2500 nm," *Journal of Biomedical Optics*, 4(1):36-46, 1 Jan 1999. doi:10.1117/1.429919

[11] Lemaillet, P, J-P Bouchard, J Hwang, and DW Allen: "Double-Integrating-Sphere System at the National Institute of Standards and Technology in Support of Measurement Standards for the Determination of Optical Properties of Tissue-Mimicking Phantoms," *Journal of Biomedical Optics*, 20(12):1-8, 27 Oct 2015. doi:10.1117/1.JBO.20.12.121310

[12] Lemaillet, P, CC Cooksey, ZH Levine, AL Pintar, J Hwang, and DW Allen: "National Institute of Standards and Technology Measurement Service of the Optical Properties of Biomedical Phantoms: Current Status," *Proc SPIE Int Soc Opt Eng*, 9700, 24 Mar 2016. doi:10.1117/12.2214569

[13] Hanssen, LM and KA Snail: "Integrating Spheres for Mid-and near-Infrared Reflection Spectroscopy," in *Integrating Spheres for Mid-and near-Infrared Reflection Spectroscopy*, 2, Peter Griffiths and John M. Chalmers Eds. Chichester:John Wiley & Sons, pp.1175-1192 15 Jun 2002, ISBN: 9780470027325. doi:10.1002/0470027320 See also NIST 841543

[14] Hamamatsu. (10 Nov 2022). *16-Element Si Photodiode Arrays*.ed. [data sheet]. https://www.hamamatsu.com

[15] NMks/Newport. (11 Nov 2022). *Photomultiplier Tube, 160-900 nm Range, Quartz Window 77348*.ed. Products/Light Analysis/Optical Sensors/Low Light Sensors/Photomultiplier Tubes/. https://www.newport.com/p/77348

[16] Zijlstra, WG, A Buursma, and OW van Assendelft, *Visible and near Infrared Absorption Spectra of Human and Animal Haemoglobin Determination and Application*, 1st ed. London:CRC Press, pp. 28 Apr 2000 ISBN:9780367447403. doi:10.1201/9780429071096

[17] Beckman, AO, WS Gallaway, W Kaye, and WF Ulrich: "History of Spectrophotometry at Beckman Instruments, Inc," *Analytical Chemistry*, 49(3):280A-300A, Mar 1977. doi:10.1021/ac50011a710

[18] Simoni, RD, RL Hill, M Vaughan, and H Tabor: "A Classic Instrument: The Beckman DU Spectrophotometer and Its Inventor, Arnold O. Beckman," *Journal of Biological Chemistry*, 278(49):79-81, 5 Dec 2003. doi:10.1016/S0021-9258(20)75750-9

[19] Hewlett Packard: "Compound Identification with Hp 8450 A Uv Visible Spectrophotometer," *Analytical Chemistry*, 51(12):1188A-1189A, Oct 1979. doi:10.1021/ac50048a728